\documentclass[trackchanges,twocolumn]{aastex7}

\usepackage{colortbl}
\accepted{to Research in Astronomy and Astrophysics, 2026}

\begin{document}
\begin{flushright} 
\end{flushright}
\title{Energy-independent X-Ray Polarization in 4U 1630--47 in the Intermediate State from Spectroscopy and Polarimetry} 

\author[orcid=0000-0001-5129-1491,gname=Seifina, sname='Elena']{Elena Seifina} 
\affiliation{Lomonosov Moscow State University/Sternberg Astronomical Institute,
Universitetsky Prospect 13, Moscow, 119992, Russia \email{seif@sai.msu.ru}}
\email{seif@sai.msu.ru}

\author[orcid=0000-0000-0000-0001,sname='Lev']{Lev Titarchuk}
\affiliation{INAF-IAPS Via Fosso Del Cavaliere 100, 00133, Rome, Italy}
\email{lev.titarchuk@inaf.it}
\affiliation{251 Rue Bernard Blier Tartugui 84400 GARGAS France}

\author[gname=Paolo,sname=Soffitta]{Paolo Soffiitta}
\affiliation{INAF-IAPS Via Fosso Del Cavaliere 100, 00133, Rome, Italy}
\email{paolo.soffitta@inaf.it}

\author[gname=Lorenzo,sname=Marra]{Lorenzo Marra}
\affiliation{INAF-IAPS Via Fosso Del Cavaliere 100, 00133, Rome, Italy}
\email{lorenzo.marra@inaf.it}

\author[gname=Fiamma,sname=Capitanio]{Fiamma Capitanio}
\affiliation{INAF-IAPS Via Fosso Del Cavaliere 100, 00133, Rome, Italy}
\email{fiamma.capitanio@inaf.it}

\author[gname=Fabio,sname=Muleri]{Fabio Muleri}
\affiliation{INAF-IAPS Via Fosso Del Cavaliere 100, 00133, Rome, Italy}
\email{fabio.muleri@inaf.it}



\author{Anastasiia Emelianova}
\affiliation{
Lomonosov Moscow State University, Faculty of Physics, GSP-1, Moscow, Leninskie Gory, 1 
119991 Russia \email{hastysa02@gmail.com}}
\email{hastysa02@gmail.com}





\begin{abstract}
The IXPE mission enabled measurements of X-ray polarization in accreting binaries.  The dependence of the X-ray polarization degree  on the binary inclination  and the 
optical depth of the emitting medium $\tau_0$  are the new  important tools  in X-ray Astronomy. We analyzed  the black hole X-ray binary 4U~1630--47 as an example of  one of the enigmatic sources in our Galaxy.  We consider  spectral and polarimetric observations of outburst episodes from 4U~1630--47 observed with the IXPE and the  NICER. We carefully  investigated a time interval of the IXPE observations during which 4U~1630--47 was in a single spectral state (the intermediate  state) and  found that the polarization degree
 did not change significantly during this time interval   with the photon energy in the 2--8 keV range ($P\sim$ 6\%). We interpreted this energy independence behavior  in $P(E)$ as a result of the Comptonization (up-scattering) of soft photons in the flat  Compton cloud. We argued that the X-ray source broadband energy spectra  can be reproduced by a physical model composed of the Comptonized component and  iron-line ({\it Gaussian}) component. Spectral analysis using this model shows that during the intermediate state the 4U~1630--47 spectrum was characterized by a photon index  $\Gamma\sim 2.5$ and the  electron temperature $kT_e\sim ~9 ~{\rm keV}$. In fact, this approach allows one to easily find the half-optical depth $\tau_0\sim 1.8$ of the Compton cloud in 4U~1630--47 applying the inclination $69\pm1^\circ$ and using the relation $(P,\cos(i),\tau_0)$ when solving the full radiative transfer equation for a flat Compton cloud by an iterative method. This approach is also accompanied by the fact that the polarization degree is independent of the photon energy, which we demonstrate observationally, at least for 4U~1630--47 in the 2--8 keV range. 
\end{abstract}




\section{Introduction}           
\label{sect:intro}

Determining the orbital inclination ($i$) of X-ray binaries (XRBs) is crucial for accurately estimating the masses of the compact object ($M_1$) and the companion star ($M_2$). This requires modeling ellipsoidal variations in optical light curves or X-ray eclipse data constrained by a mass function $f(M)$. 
In turn, this estimate requires knowledge of the binary period, the component mass ratio ($q$), and measurements of the radial velocity of the companion star [see \cite{Orosz03} and references therein].  
The absence of X-ray eclipses and rotational broadening of absorption lines (radial velocity shifts) in the stellar optical spectra also hinders these estimates. Furthermore, there are many low-mass binaries for which optical emission is inaccessible due to galactic absorption.
Finally, analyzing the optical emission of binaries to determine the masses of their components is only possible in those binaries where accretion ceases and the binary returns to a quiescent state. 
In contrast, based on X-ray observations and analysis of X-ray timing characteristics of binaries, it is possible to estimate the orbital inclination of X-ray binaries even when the object is in an outburst  state, making traditional dynamical methods inapplicable.

4U~1630--47 is a low-mass X-ray binary system with a black hole (BH), which attracts astronomers with its frequent X-ray outbursts. This BH is relatively poorly studied due to its location in the galactic plane, which obscures it with dense clouds of gas and dust. Therefore, many of its parameters, such as the BH mass and binary inclination, are still the subject of scientific debate. In particular, the inclination of 4U~1630--47 is estimated only by indirect methods. Specifically, it is calculated by scaling spectral and timing characteristics [e.g., \cite{RodriguezCavero23}] using the X-ray binaries XTE~J1550--564, GRO~1655--40, and H~1743--322 as reference sources and assuming a distance $>$10 kpc \citep{STS14}. Based on the detection of short--duration dips in the X-ray light curve of 4U 1630-47 during the outburst \cite{Kuulkers98} also  inferred a relatively high inclination of 60$^{\circ}$ -- 75$^{\circ}$. Although this inclination is generally accepted given certain spectral characteristics, the precise inclination of 4U~1630--47 is a subject of scientific debate. Optical observations of the companion star in this source to refine the inclination are controversial, as the system lies in the Galactic plane and is obscured by dense clouds of gas and dust. Since the inclination directly affects the BH mass estimate, it is worth considering alternative approaches to determining the tilt of 4U~1630--47 using X-ray observations. 

A new X-ray method for estimating the binary inclination was recently developed by \cite{Tetal25}, hereafter Tetal25. This method relies on measuring  the degree of the  linear polarization (LP) in sources with accreting BHs taking into account its energy spectral characteristics. Tetal25 discovered that X-ray emission from accretion disks is inclination-sensitive. According to theoretical estimates of \cite{ST85}, hereafter ST85, this emission should be linearly polarized if the Compton Cloud (CC, corona) has a ``flat pancake'' shape. This polarization results from the multiple scattering of initially soft photons in a hot, flat, optically dense CC. ST85 suggested that the LP degree of this emission depends on the inner disk (CC) inclination to the line of sight and the optical depth of the flat CC. The optical depth of CC can be easily estimated using the photon index $\Gamma$ and the plasma temperature $kT_e$ from spectral analysis of the source's emission. 
Recently, the Tetal25 method for estimating the inclination $i$ was successfully tested using the BH XRB 4U~1957+115 based on IXPE, NuSTAR, NICER, {\it RXTE}, Swift, Suzaku, and ASCA data \citep{TSSM25}.

The object 4U~1630--47 was recently observed by the IXPE space observatory, launched on December 9, 2021, which, for the first time, allowed  a sensitive X-ray polarimetric study of sources of this class [see \cite{soffitta21}, \cite{Weisskopf22}]. \cite{Ratheesh24} detailed a possible interpretation of the polarization properties of 4U~1630--47 using IXPE data and timing variations in the source's polarization properties (see Figure 4 there). \cite{Ratheesh24} argued that the polarization degree increases with energy as a result of the combined contributions of the accretion disk, disk wind, and weak corona.

Using polarimetric observations of 4U~1630--47 with the IXPE, \cite{Kushwaha23} (hereinafter K23) proceeded  with a detailed analysis of this source during its outburst in 2022. They found linear polarization degree (PD) $P = 8.3 \pm 0.2 $ \% in the $2-8$ keV range.  
A linear polarized phenomenon, as they proposed,  is a result   of  combined effects an  accretion disk (AD), disk wind, and a weak corona.  They even suggested  the spin  of the BH in 4U~1630--47   estimated  with an application   of  the  IXPE data.  K23 used the XSPEC relativistic model {\it  kerrbb} to fit the available  IXPE data. 

Furthermore, ST85 showed that polarized X-ray emission from accreting  sources  with a flat geometry has a polarization independence of a photon energy (at least in the non-relativistic X-ray energy range, below 10 keV). We intend to test this interesting theoretical conclusion by ST85 for 4U~1630--47.

\begin{figure*}
\centering
\includegraphics[scale=0.87, angle=0]{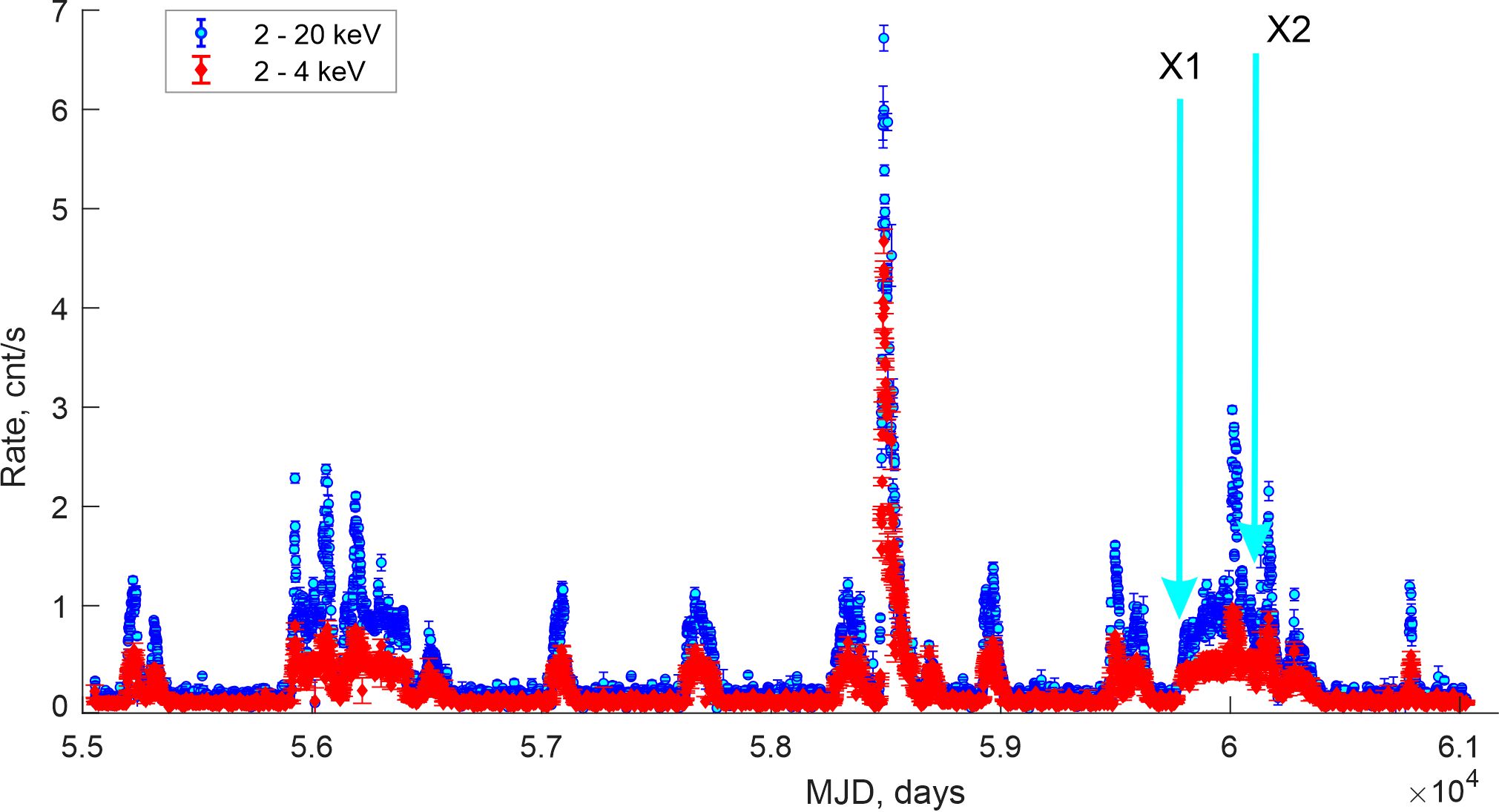} 
\caption{Evolution of MAXI count rate during 2009--2025 (MJD 55053--61027) 
observations of 4U~1630--47. The observation times $X1$ and $X2$ are indicated by blue arrows (see Table~\ref{tab:table_Nicer+IXPE}).
}
\label{lc_1}
\end{figure*}


\begin{figure*}
\centering
\includegraphics[scale=0.78, angle=0]{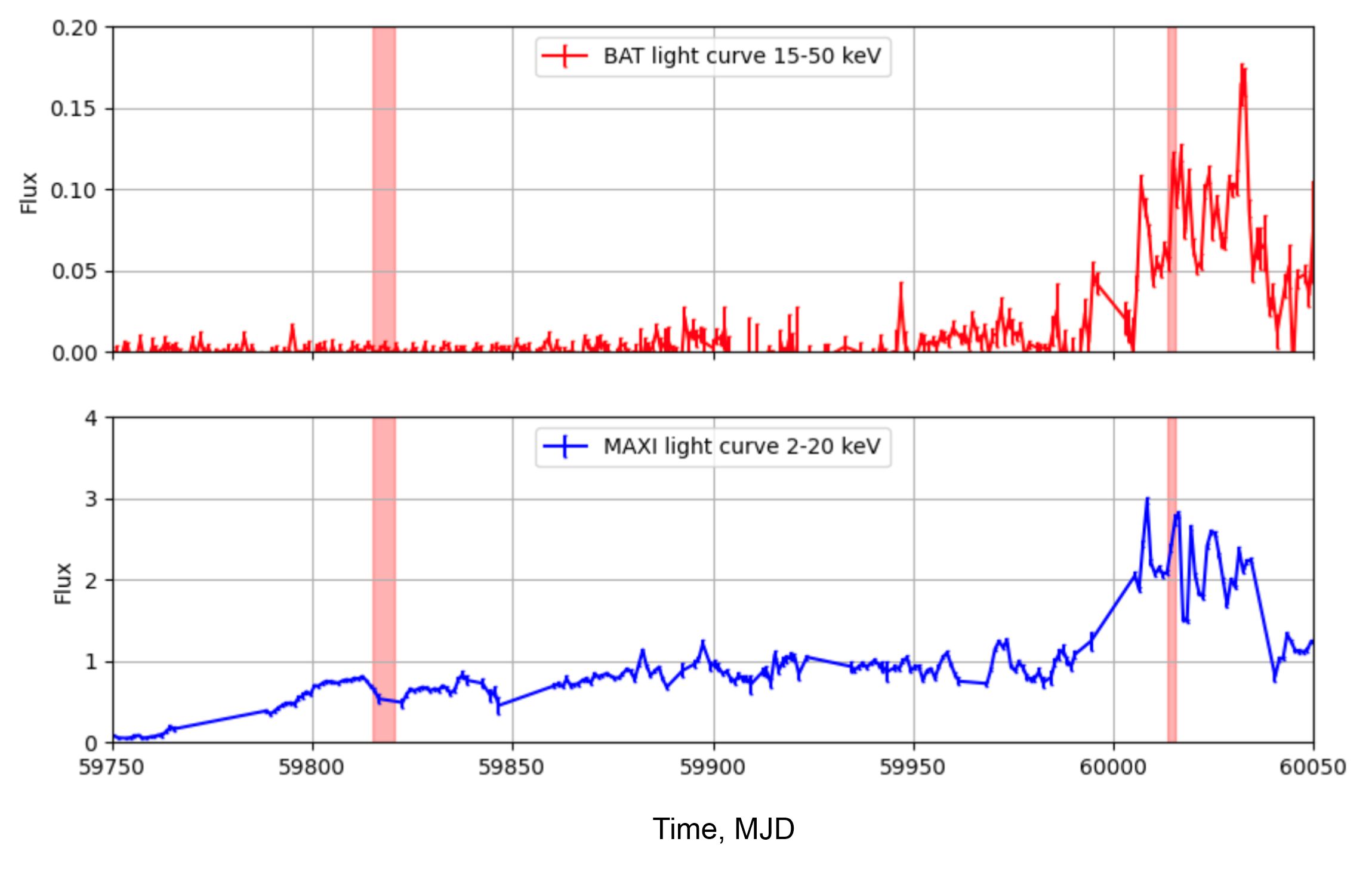} 
\caption{Evolution of MAXI count rate during 2022 -- 2023 (MJD 59750--60050) 
observations of 4U~1630--47.
}
\label{lc_2}
\end{figure*}

%
%

\begin{table*}
 \centering 
 \caption{Details of  the {\it NICER} and  the {\it IXPE} observations of 4U~1630--47. }
   \label{tab:table_Nicer+IXPE}
 \begin{tabular}{llllcr}
 \hline\hline  
Epoch & Mission & Obs. ID& Start time (UT)  & Exposure (s) & MJD interval \\
      \hline
Nc1 & {\it NICER}  & 5501010102  & 2022 Aug 23 00:24:42 & 1950&59814.0--59814.9$^{1,2}$ \\
 Nc2 & {\it NICER}  & 5501010103  & 2022 Sep 7 06:00:41 & 452 &59829.2--59829.3$^{1,2}$ \\
Nc3 & {\it NICER}  & 5501010104  & 2022 Aug 25 01:42:36 & 3920&59816.0--59816.2$^{1,2}$ \\
\hline 
X1 & {\it IXPE}  & 1250401 &  2022 Aug 23 23:15:32 & 458518 & 59814.9--59825.8$^{1,3,4}$\\
X2 & {\it IXPE}  & 2250601   &  2023 Mar 10 18:47:59 & 137717 & 60013.8--60016.7$^{3,4}$ \\
\hline
      \end{tabular}
\\
(1) \citet{Kushwaha23};
(2) \citet{Rawat23};
(3) \citet{Ratheesh24}; and 
(4) \citet{RodriguezCavero23}. 
\end{table*}

%
%

\begin{table}
 \caption{Basic information on 4U 1630--47}
 \label{tab:parameters_binaries}
 \centering 
 \begin{tabular}{ll}
 \hline                      
Source parameter               &     Value \\
\hline
Mass of primary star M$_1$ (M$_{\odot}$)   & 10.0$\pm$0.1$^{~(a)}$ \\
Class of primary star &  BH$^{~(a,b)}$ \\
Mass of secondary star M$_2$ (M$_{\odot}$)  & 2--20$^{~(b)}$ \\
Class of secondary star & Be-star$^{~(b)}$ \\
Inclination,  $i$,  deg           & $\sim$70$^{~(a)}$, $i \leq 73^{~(e)}$                         \\
Polarization degree, \% & 6.86$\pm$0.22 (HSS)$^{~(c,d)}$, \\
 &  8.44$\pm$0.18 (LHS)$^{~(c,d)}$ \\
{Polarization angle, deg} & {21.3$\pm$0.9 (HSS)$^{~(h)}$,} \\
  & {$\sim$18 (LHS)$^{~(h)}$} \\
Distance, $d$,  kpc       & 10--11$^{(a,b)}$ \\
RA (J2000), $\alpha$       & 16$^h$ 34$^m$ 01.61$^{s~(e,f,g)}$  \\
Dec (J2000), $\delta$       & --47$^{\circ}$ 23{\tt '} 34.8{\tt ''}$^{~(e,f,g)}$ \\
{Inner and outer radii of the annular} & {130 and 252 }\\
{background extraction area, {\tt ''}} & \\
{Outburst activity period, $P$,  days}       & {$\sim$600$^{(a,b,j)}$} \\
Alternative names &  4U 1630--472, \\
                          &SWIFT J1633.8--4724A\\
$N_{H,Gal}$, cm$^{-2}$   &(1--2)$\times$10$^{22~(a)}$  \\
$K$, mag        & $\approx$16--18$^{~(b)}$  \\
 \hline                                             
 \end{tabular}
\\$^{(a)}$ \cite{STS14};
$^{(b)}$  \cite{Augustejin01}; 
$^{(c)}$ \cite{Rawat23,Kushwaha23}; 
$^{(d)}$ \cite{Ratheesh24,RodriguezCavero23};
$^{(e)}$ \cite{Hjellming+Kuulkers98};
$^{(f)}$  \cite{Buxton98};
$^{(g)}$  \citep{Hjellming99}; 
{$^{(h)}$} \cite{RodriguezCavero23};
{$^{(i)}$} \cite{Das25} {and}  
{$^{(j)}$} \cite{Capitanio15}. 
\end{table}

We present our analysis of available {\it IXPE}, and  {\it NICER} observations of 4U~1630--47. In Section~\ref{sect:data} 
we show the list of observations used in our data analysis, and 
we provide the details of our data analysis, 
a description of the spectral models used for fitting these data and  the corresponding results in section~\ref{sect:results}. 
In section~\ref{sect:results_discussion} 
we discuss the main results of the paper and  produce our conclusions.

%
%

\begin{figure*}
\centering
\includegraphics[scale=1.2,angle=0]{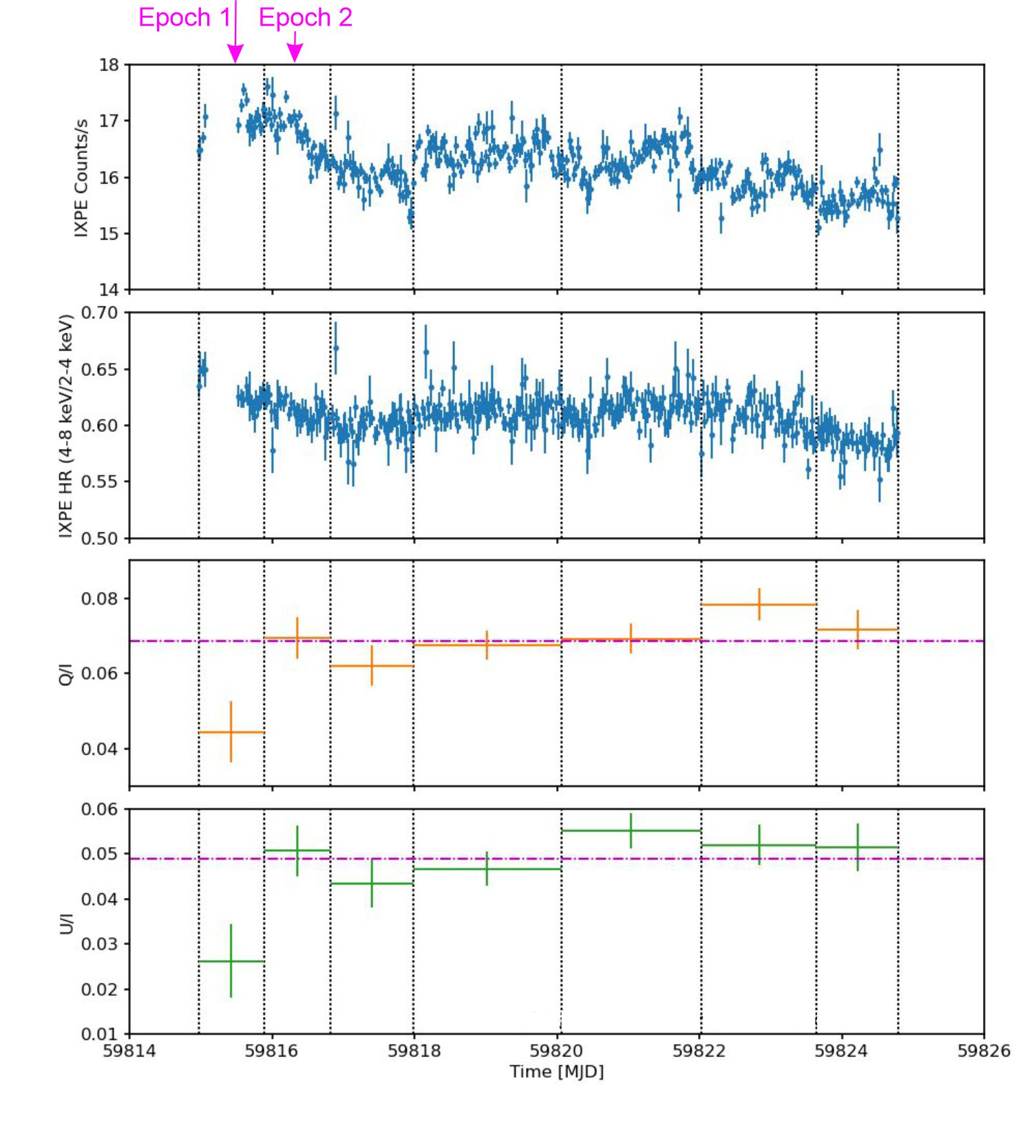}
\caption{Light curve 
 and time-dependent polarization properties of 4U~1630--47 from the IXPE observation 
 $X1$.
 IXPE light curve in the 2--8 keV energy range with 800 s time bins (top); time evolution of the hardness ratio, defined as the ratio of IXPE count rates in the 4--8 keV and 2--4 keV bands, using the same 800 s time bins (middle); and normalized Stokes Q and U parameters, measured by the IXPE as a function of time (two bottom panels). Dashed lines show the best-fit model. The horizontal dashed line represent the best fit of the normalized Q and U parameters with a constant value. 
{Epoch 1 and Epoch 2, analyzed in detail in this paper, are indicated by arrows at the top of the figure.}
 }
\label{fraq_1630}
\end{figure*}

%
%

\begin{figure*}
\centering
\includegraphics[scale=1.2,angle=0]{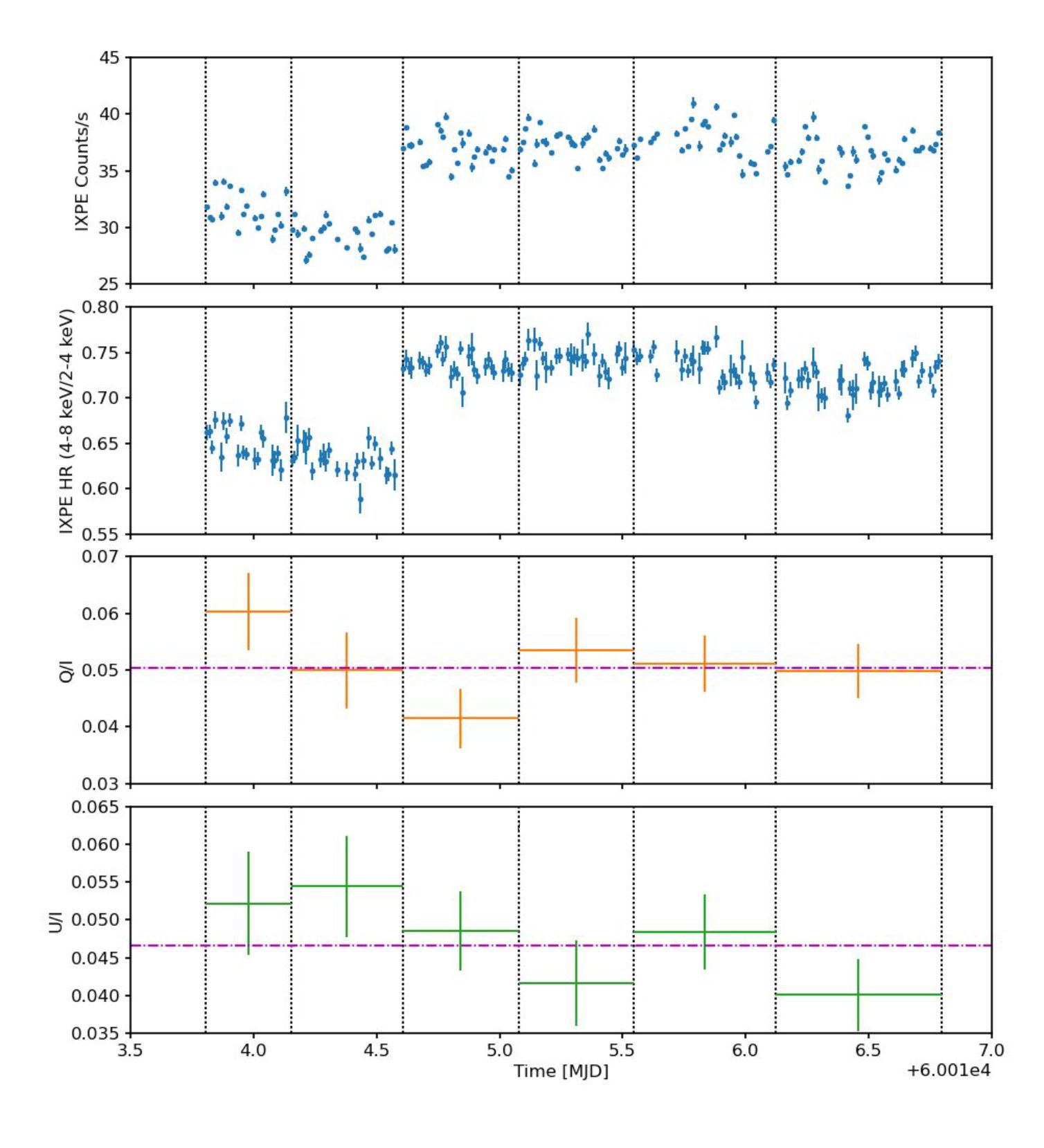} 
\caption{
{The same as in Fig.~\ref{fraq_1630}, for observation $X2$.}
 }
\label{fraq_1630_b}
\end{figure*}

%
%

\begin{figure*}
\centering
\includegraphics[scale=0.8,angle=0]{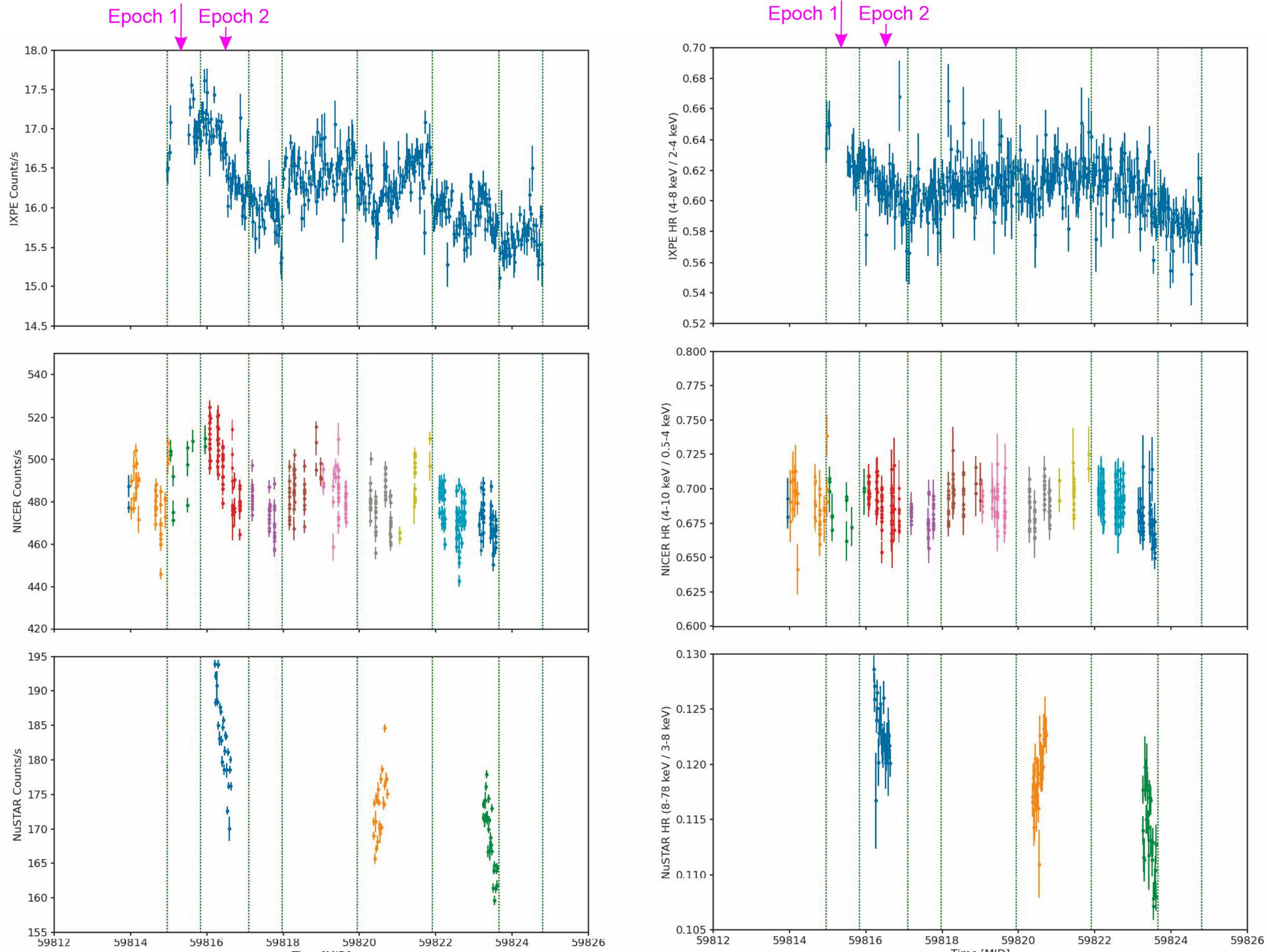} 
\caption{
{Comparison of 4U~1630--47 properties: count rate (left column) and hardness ratio (right column) from quasi-simultaneous observations of IXPE ($X1$, top panels), NICER (middle panels), and NuSTAR (bottom panels). Epoch 1 and Epoch 2, analyzed in detail in this paper, are indicated by arrows at the top of the figure.}
 }
\label{lc+HR_IXPE_NUSTAR_NICER}
\end{figure*}

%
%
\begin{figure*}
\centering
\includegraphics[scale=0.90,angle=0]{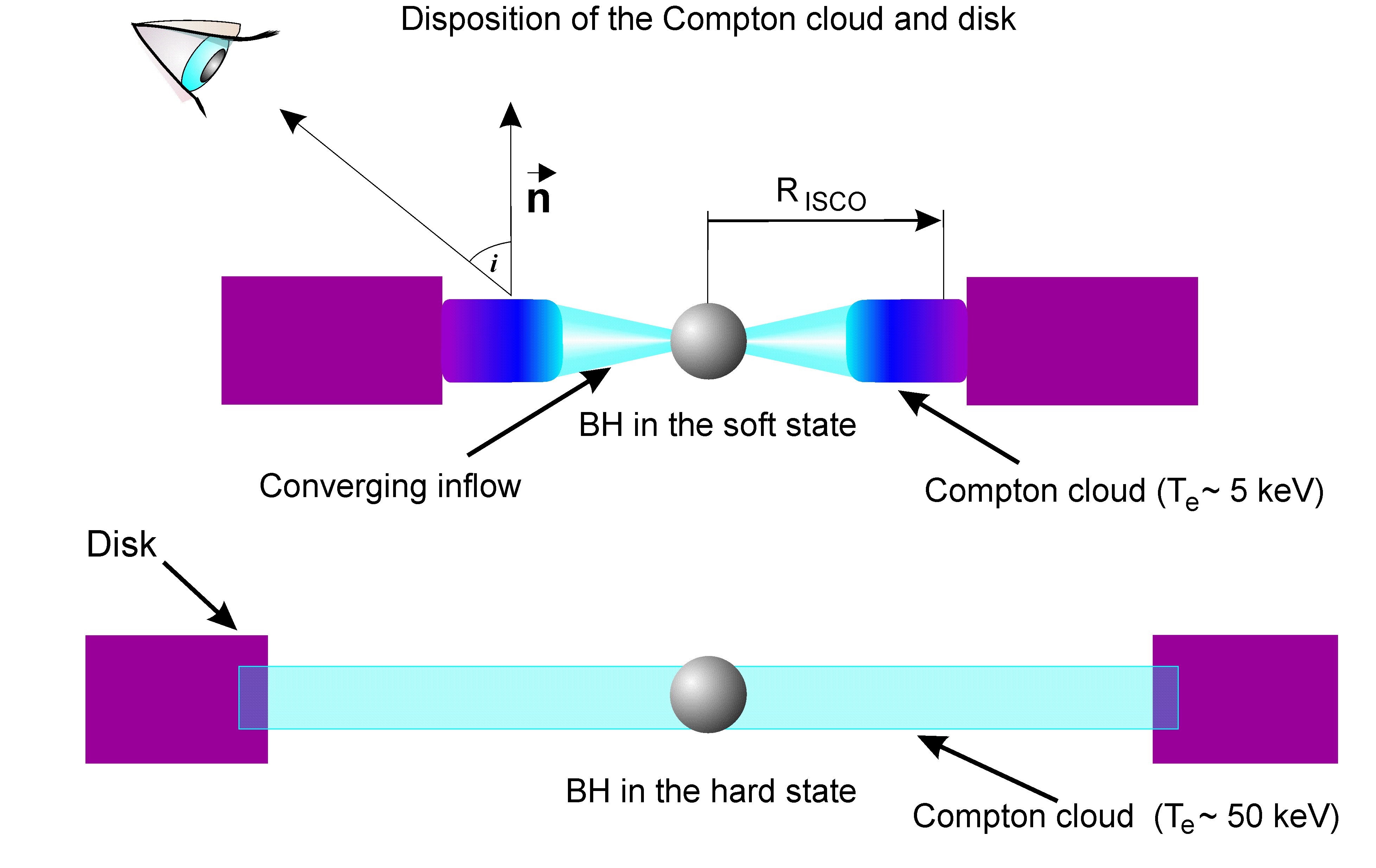} 
\caption{Model of a BH with the Compton cloud and an accretion disk
 }
\label{model
}\end{figure*}

\section{Data selection }
\label{sect:data}
We used observations of 4U~1630--47  by {\it NICER} (April 2022) and {\it IXPE} (August 2022 and March 2023; see Table~\ref{tab:table_Nicer+IXPE}). We extracted these data from the HEASARC archives and found that these data  cover a wide range of X-ray luminosities (see Fig.~\ref{lc_1}). The {\it  NICER} and {\it IXPE} data do not strictly coincide in time, so we examined in more detail the 4U~1630--47 data obtained simultaneously with the {\it IXPE} observations at Epoch $X1$ (see Table~\ref{tab:table_Nicer+IXPE}). Although the low photon statistics did not allow us to obtain any information about the state of the source, we analyzed data using  {\it  NICER}  observations located nearby.

\subsection{{\it NICER} data\label{nicer data}}

{\it NICER} observed 4U~1630--47 on 23 August 2022 between MJD~59814.0 and MJD~59814.9, shortly before the {\it IXPE} campaign (see Table~\ref{tab:table_Nicer+IXPE}). In addition to the $Nc1$ observation associated with Epoch~$X1$, we also analyzed the $Nc2$ and $Nc3$ observations to investigate the evolution of the spectral properties of 4U~1630--47 during the subsequent stages of the August 2022 outburst. 

Data reduction was carried out using the {\tt NICERDAS} software package together with the latest calibration database ({\tt CALDB}). Standard calibration and screening procedures were applied with the {\tt nicerl2} task to produce cleaned event files. Source and background spectra, as well as the corresponding response matrices, were generated in the 0.5--12~keV energy range using the {\tt nicerl3-spect} task. Finally, the spectra were grouped to ensure a minimum of 20 counts per energy bin for the spectral fitting. 

\subsection{  The {\it IXPE} data\label{ixpe data}}
The available {\it IXPE} observations of 4U~1630--47 were obtained on 23 August 2022 (ObsID~01250401) and 10 March 2023 (ObsID~02250601), with net exposure times of 458.5~ks and 137~ks, respectively. 

Level-2 cleaned event files were used to extract source products for both spectroscopic and polarimetric analyzes in the 2--8~keV energy range. Source events were selected from a circular region centered on the source position with a radius of 120\arcsec. The extraction and filtering procedures were carried out using the {\tt XSELECT} software, applying effective event-number weighting to the Stokes parameters $I$, $Q$, and $U$. 

Background events were extracted from an annular region centered on the source, with inner and outer radii of 130\arcsec\ and 252\arcsec, respectively (see Table~\ref{tab:parameters_binaries}). The data products were generated in the form of ``polarization cubes,'' allowing flexible slicing by detector unit, time interval, and energy range. 

Polarization cubes were produced using the unweighted {\tt pcube} algorithm implemented in the {\tt IXPEOBSSIM} package \citep{Baldini22}, allowing a model-independent analysis of polarimetric data.

{Note that due to the IXPE X-ray observatory's near-equatorward, circular orbit at an altitude of approximately 600 km with an orbital period of 96.6 minutes, periodic dips and interruptions in the light curves are observed. These are caused by Earth occultations, when the source disappears below the planet's horizon, as well as by the satellite's passage through the South Atlantic Anomaly.}

{It should be noted that during both IXPE epochs, NICER, and NuSTAR also conducted simultaneous observations of 4U~1630--47 on March 10, 2023. However, the NuSTAR observations were not used because they do not fall within Epoch 1. 
In fact, as can be seen from Fig.~\ref{lc+HR_IXPE_NUSTAR_NICER}, the NuSTAR observations begin in Epoch 2. Therefore, in our analysis for epoch 1, the NuSTAR observations will not be informative for the purposes of this paper.}

%
%

\begin{figure*}
\centering
\includegraphics[scale=1.1,angle=0]{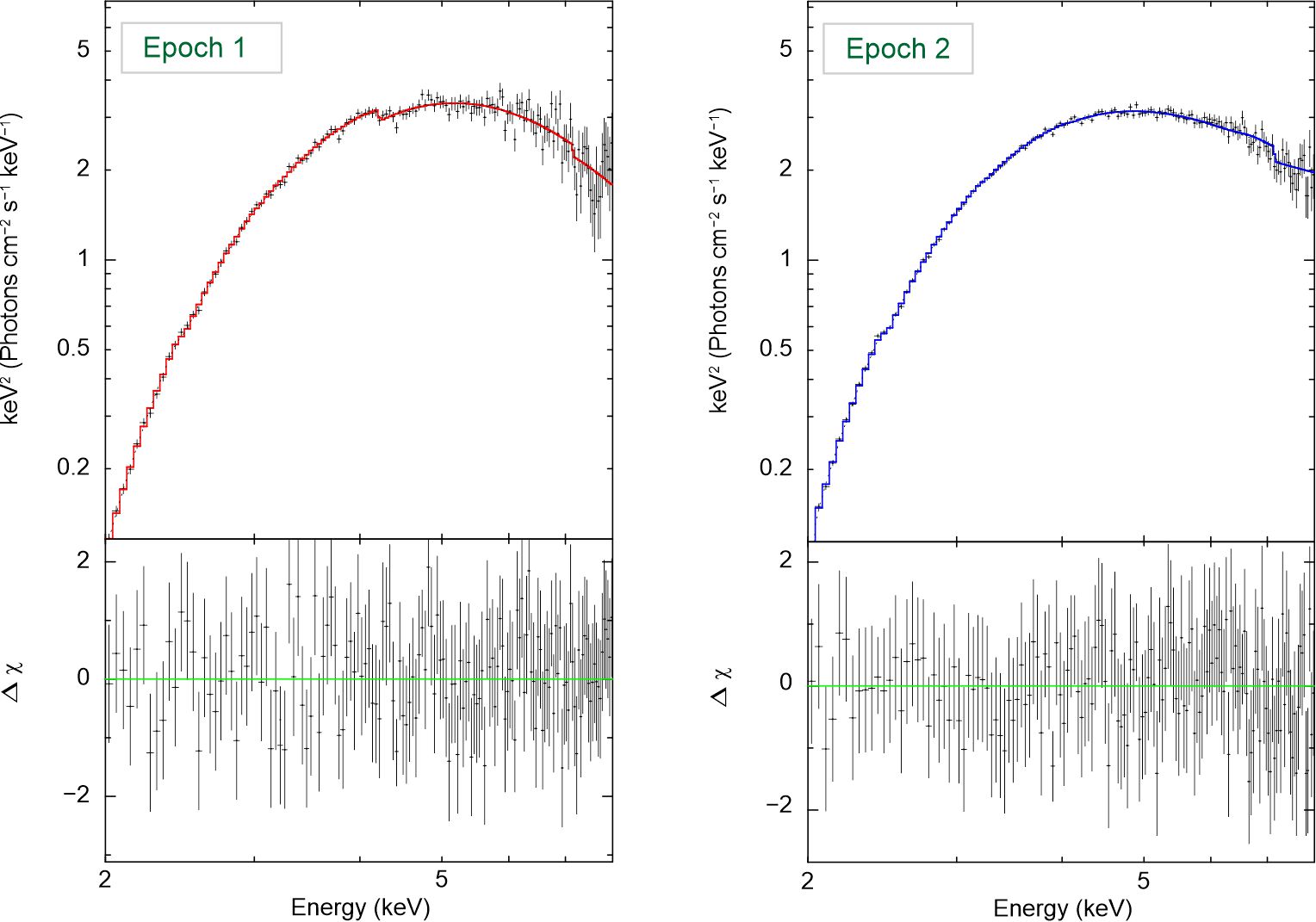} 
\caption{X-ray spectrum of 4U~1630--47 observed by the {\it IXPE} (ID=01250401). The data are shown by black crosses, and the spectral model is displayed by red line {for {\it Epoch 1} and blue line for {\it Epoch 2}}. Bottom panels: $\Delta\chi$ versus photon energy in keV. 
 }
\label{sp_ixpe}
\end{figure*}


\section{Results }
\label{sect:results}
\subsection{Light Curve Analysis \label{lc analysis}}

Object 4U~1630--47 exhibits X-ray outbursts with a period of approximately 600 days (Fig. ~\ref{lc_1}). We studied the X-ray light curve of 4U~1630--47 to determine the observation time interval during which the X-ray source was strictly in one of the spectral states [Low-Hard State (LHS), Intermediate State (IS), or High-Soft State (HSS)] and did not make any transitions between them. To do this, the light curve for both observations ($X1$ and $X2$) was divided into time segments, as shown in Fig.~\ref{fraq_1630}.
{When dividing the segments, we used the following criteria: approximately equal countrate and hardness ratio within a 1--2-day segment. Our division also ensured sufficient statistical significance, for example, for the dependence of PD on energy for different intervals.} We estimated the spectral hardness and degree of polarization for each segment and found that these characteristics vary from segment to segment. 
{We mainly focus on the segment of $X1$-observation (ID=01250401) from August 23, 2022 23:15:32.1 to August 24, 2022 19:36:19.7 (MJD=59814.9 -- 59815.8, hereinafter -- {\it Epoch 1}).}  
{Additionally, we also study in detail the second interval of the same IXPE observation from August 24, 2022, 20:08:45.6 to August 26, 2022, 02:07:59.2 (MJD=59815.8 -- 59816.8, hereinafter referred to as {\it Epoch 2})  for a test comparison of source polarization properties with those in {\it Epoch 1} (see Figs.~\ref{fraq_1630} and \ref{lc+HR_IXPE_NUSTAR_NICER}).
}

In the analysis of the soft-state observation of 4U~1630--47, \cite{Ratheesh24} reported a time-dependent behavior of the degree of polarization during the {\it IXPE} observation. In particular, they found a lower degree of polarization during the initial part of the observation $X1$ (see Figure 4 there). Motivated by this result, we further investigated the timing variability of the polarization properties by dividing the observation into different time intervals. 

The results of this analysis are shown in Figure \ref{fraq_1630}. By fitting the normalized Q and U Stokes parameters with a constant value, we find that during the $X1$ observation, the initial time segment exhibits polarization properties that clearly differ from those observed during the remainder of the observation ($\chi^2_{dof}$ = 28.22/12). Specifically, for the {\it Epoch 1} segment in the $X1$ observation, the light curve shows a constant count rate, which we interpreted as indicating that the object was in a specific spectral state. For the second observation ($X2$), the constant fit cannot be rejected, indicating that there is no significant variability and therefore consistent polarization properties throughout the observation.

The presence of an object in a fixed spectral state is necessary to test the theoretical conclusion by ST85 that the degree of polarization is independent of the photon energy. Therefore, the {\it Epoch 1} segment we selected is extremely suitable for the purposes of our paper on verifying the energy-independence of the polarization degree of X-rays.

%
%
\begin{figure*}
\centering
\includegraphics[scale=0.86,angle=0]{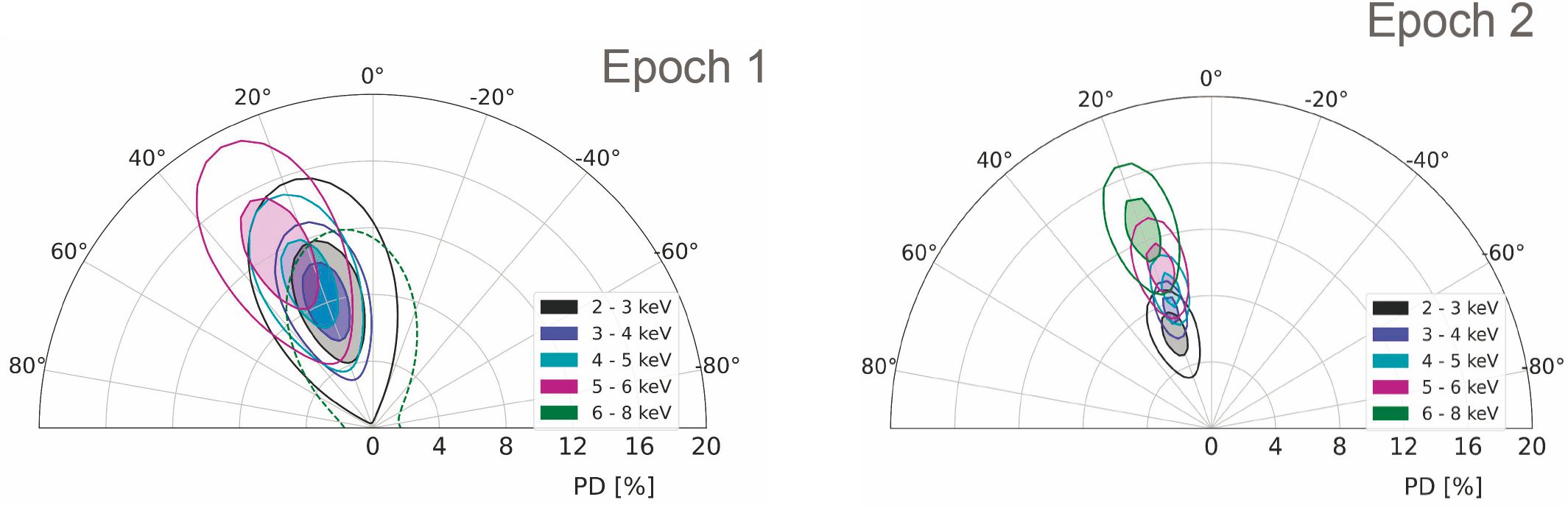} 
\caption{
{Energy-dependent X-ray polarization of 4U~1630--47. Polarization degree and polarization angle, derived from the IXPE observations $Epoch 1$ (left) and $Epoch 2$ (right), in five energy bands (labeled in different colours). 
The shaded and unshaded ellipses indicate the 68.3\% and 99.7\% confidence contours, respectively.
}
}
\label{PD_E}
\end{figure*}

%
%
\begin{figure*}
\centering
\includegraphics[scale=0.8,angle=0]{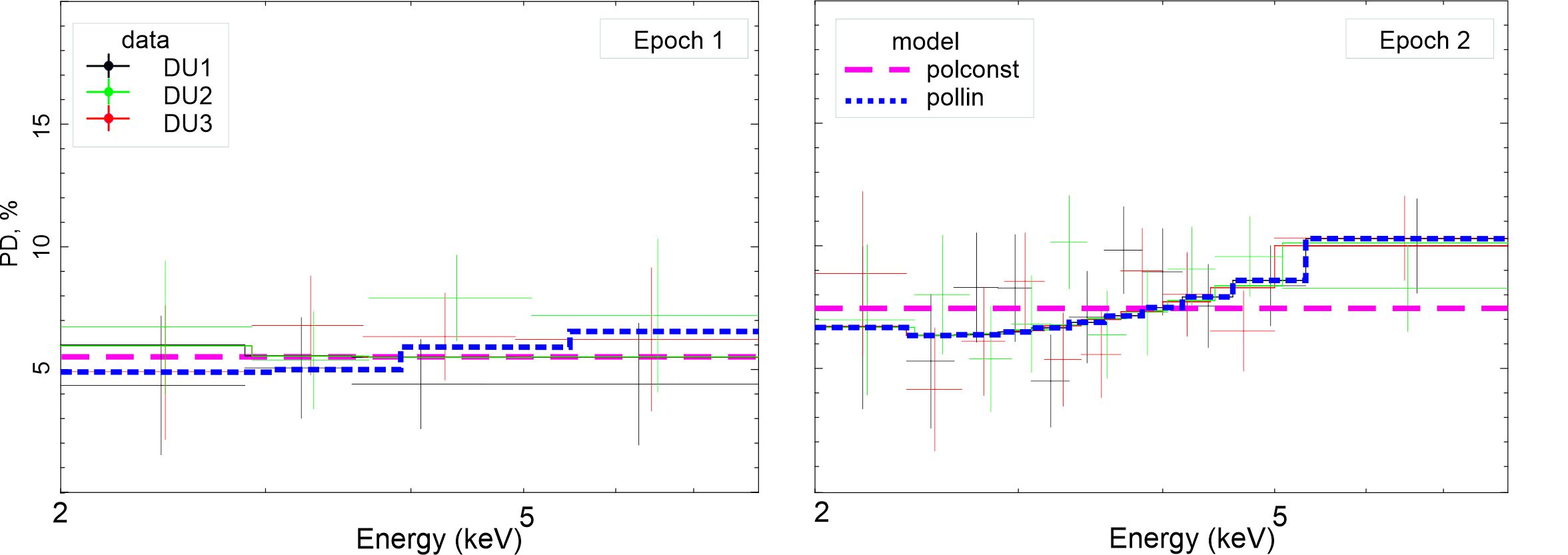} 
\caption{The polarization degree 
{PD} versus the photon energy 
using the IXPE observation $X1$ (ID=1250401) for MJD 59814.9 -- 59815.8 
{({\it Epoch 1}, left panel) and MJD 59815.8 -- 59816.8 ({\it Epoch 2}, right panel)}. {Pink} dashed line indicates the mean value of the  polarization degree 
{with Model 1}, while {blue dotted line marks PD(E) with Model 2}. Error bars correspond to 68\% confidence level.
}
\label{PD_E}
\end{figure*}

\subsection{Spectral Analysis \label{spectral analysis}}
The broadband X-ray spectrum of 4U~1630--47 was analyzed using the {\tt XSPEC} v12.14.1 software package. We adopted a spectral model consisting of a Comptonization continuum described by the {\tt CompTB} model \citep{ft11}, together with a {\tt Gaussian} component to account for the Fe K$_\alpha$ emission line in the 6--8~keV energy range. 

The {\tt CompTB} model is applicable to different spectral states and Comptonization regimes (thermal and/or bulk-motion Comptonization) observed in black hole and neutron star X-ray binaries \citep{TS24a,TS24b,TSC23,TS23,TS21,TSCO20,TS17,TS16,tsei16b,STS16,STS14,tss10,ts09}. In this framework, the observed continuum is produced by Compton up-scattering of soft seed photons in the Compton cloud (CC). The seed photons are assumed to originate from a blackbody-like component characterized by a color temperature $kT_s$. 

In the {\tt CompTB} formalism, the spectral normalization $N_{\rm Com}$ is proportional to the mass accretion rate $\dot{M}$, while the energy spectral index $\alpha$ (or, equivalently, the photon index $\Gamma=\alpha+1$) determines the slope of the Comptonized spectrum. The fraction of Comptonized radiation is parameterized by
\[
f = \frac{A}{1+A},
\]
where $A$ is the illumination factor. The electron temperature of the Compton cloud, $kT_e$, is another fundamental parameter of the model.

The Comptonized continuum is calculated as the convolution of the seed blackbody spectrum ($BB$) with the Comptonization Green's function $G$, which can be approximated as
\begin{eqnarray}
G(x,x_0) & \propto & x^{-\alpha}, \qquad x>x_0, \nonumber\\
G(x,x_0) & \propto & x^{\alpha+3}, \qquad x<x_0,
\end{eqnarray}
where $x=h\nu/kT_e$ and $x_0$ is the break energy at which the two asymptotic forms join. The resulting emergent spectrum can therefore be expressed as
\[
BB + f \cdot (BB * G),
\]
where $*$ denotes convolution with the Green's function. Thus, the adopted model self-consistently describes the combination of the direct blackbody emission and its Comptonized component.
\noindent where $x = hv/kT_e$, $x_0$ is the break-point of $G(x,x_0)$  where the two cases: $x>x_0$ and $x<x_0$ meet each other and the resulting spectrum is $BB+f\cdot BB*G$. Thus, the general model consists of a convolution of  a  soft BB and a Comptonized components. 
We further included three edges components in the spectral fit to model the calibration uncertainties presented in the NICER spectra at low energies. In addition to the above spectral components, we also applied a multiplicative model {\tt tbabs} to account for interstellar absorption. 

As a result, we used the {\tt {edge*edge*edge}*tbabs*(Comptb + Gauss)} model with the following parameters:
$N_H$ is the equivalent column absorption density of hydrogen $N_H$; the  energy spectral index $\alpha$, the fractions of the illumination of the CC 
by soft photons from the disk $f$ ($ = A/(1 + A)$), i.e., the relative weight of the Comptonization component; the electron temperatures CC $kT_e$ and normalizations of the BB-like component $N_{Com}$.  We also  account for the absorption edges in the 1.5--1.8, 2.5, and 6--7 keV energies, whose parameters are the centroid line energy $E_{edge}$, the depth $\tau_{max}$. As already noted, to fit the data in the 6--8 keV energy range, the above model used an additive {\tt Gaussian} component whose parameters are the line energy at the center $E_{line}$, the line width $\sigma_{line}$, and the normalization $N_{line}$. 

%
%
\begin{figure}
\centering
\includegraphics[scale=0.55,angle=0]{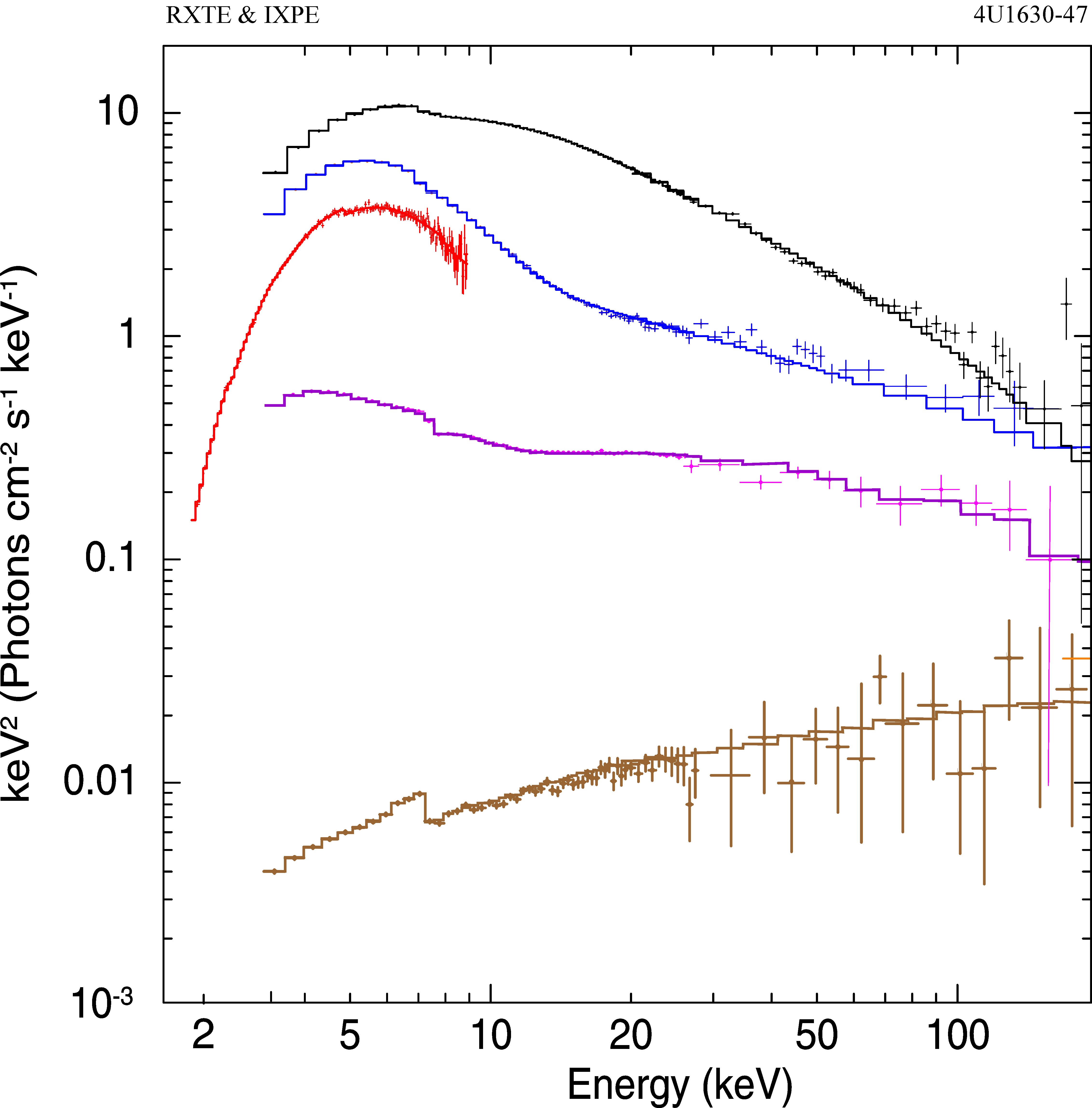} 
\caption{The IXPE spectral diagram of 4U~1630--47 in comparison with four {\it RXTE} representative source $EF_E$ spectral diagrams, related to different source spectral states. The data are taken from the IXPE observation 
{1250401, $Epoch 1$} (red, IS) and {\it RXTE} observations 30172-01-18-12 (brown, LHS),  30172-01-07-00 (violet, IS), 70417-01-03-00 (blue, IS), and 80117-01-03-00G (black, HSS). The normalization factor of 0.1 was applied for the 30172-01-18-12 spectrum for clarity.
 }
\label{IXPE_RXTE}
\end{figure}

%
%

\begin{figure*}
\centering
\includegraphics[scale=1.42, angle=0]{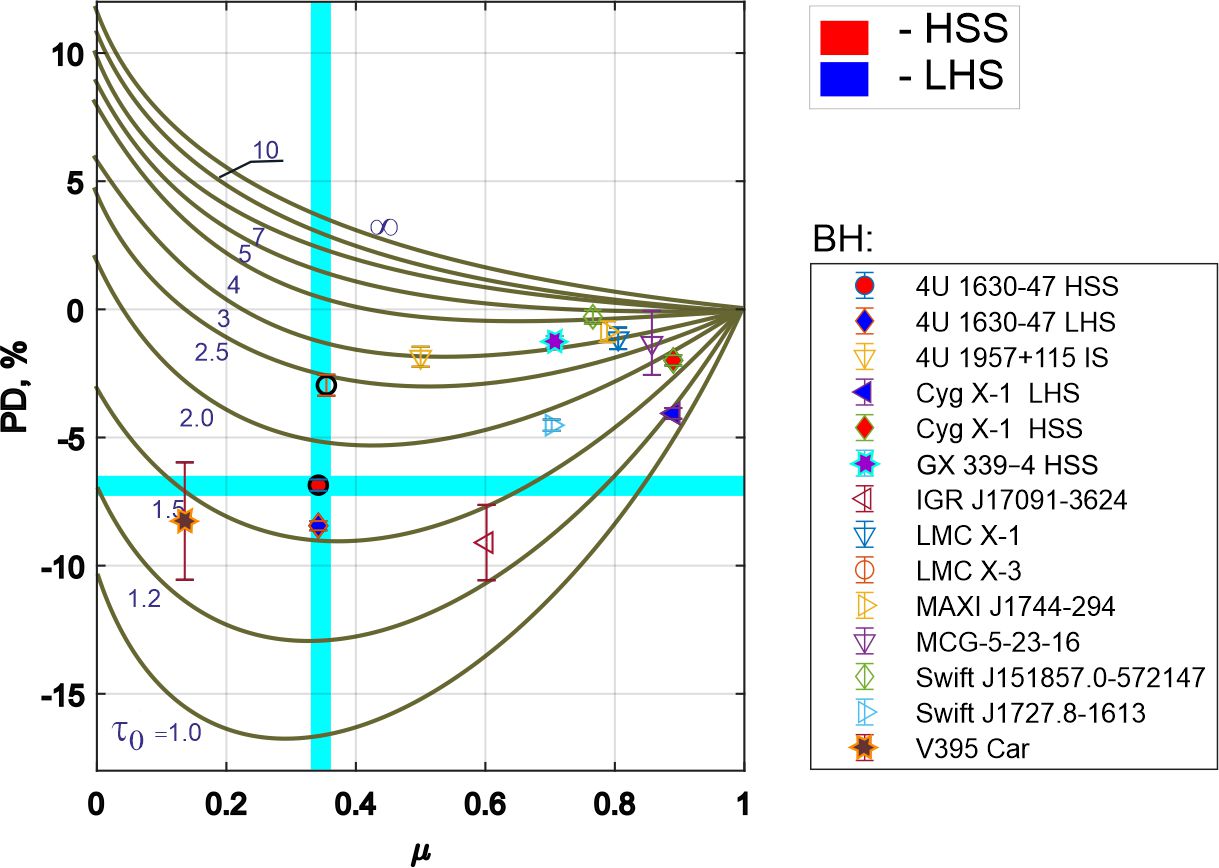} 
\caption{The linear polarization $P$ in \% using the {\it IXPE} observations (see Tetal25)  as a function of $\mu =\cos (i)$, where $i $ is the  inclination (angle between the normal to the CC 
 and the given direction to the BHs)  against a half of the optical depth of the CC 
 $\tau_0$  (taken from ST85). Red and blue fillings correspond to the soft and hard states (where acceptable). The position ($P$ and $\mu$) for 4U~1630--47 in HSS  is marked by the cyan horizontal and vertical lines. We also indicated the polarization degree for 4U~1630--47 in the LHS taken from  \cite{RodriguezCavero23}. For the remaining XRBs we used the following PD values: for 4U~1957+115 (PD=1.95$\pm$0.2\%), \cite{Marra24,Tetal25}), 
Cyg~X--1 (LHS, PD$\sim$4.0\%, \cite{Kravtsov25}; HSS, PD=(1.99$\pm$0.13\%), \cite{SteinerJames24}), GX~339--4 (PD=1.3$\pm$0.3 \%), \cite{Mastroserio25}), 
IGR~J17091--3624 (PD = 9.1$\pm$1.6\%), \cite{Ewing25}), 
LMC~X--1 (PD=1.0$\pm$0.4\%), \cite{Podgorny23}), 
LMC~X-3 (PD=3.04$\pm$0.40\%), \cite{Majumder24}),  
MAXI~J1744--294 (PD$\sim$1.3\%), \cite{Marra25}), 
MCG-5-23-16 (PD=1.08$\pm$0.66\%), \cite{Mondal24a},  
Swift J151857.0--572147 (PD=1.34$\pm$0.27\%),  \cite{Mondal24b}),  
Swift~J1727.8-1613 (PD = 4.1$\pm$0.2\%), \cite{Veledina23}), 
PD = (3.3$\pm$0.4\%) \cite{Podgorny23}) and 
V395 Car (PD = 8.5$\pm$1.6\%), \cite{Tomaru25}).
}
\label{pol_vs_mu}
\end{figure*}

\subsubsection{Spectral properties of 4U~1630--47 based on {\it IXPE} observations \label{spectral analysis_ixpe}}

During the {\it IXPE} observations on 30 August 2022, 4U~1630--47 was in a 
typical 
{intermediate} state. The best-fit parameters of the spectra in the 2--8 keV range are given in Table~\ref{tab:table_IXPE}. In particular, to reproduce the shape around  the 6.8 keV emission line, we took into account an  absorption edge at the 7.1 keV. Note, the energy of 7.1 keV corresponds to the X-ray transitions inside the iron. Here we also presented the results of a model-dependent analysis based on fitting the Stokes $I$, $Q$, and $U$ spectra of the {\it IXPE} to the XSPEC spectrum within the {\tt constant*edge*edge*edge*tbabs*polconst(CompTB+Gauss)} model 
{(Model 1)}. The observed spectrum with the best fit to the model is shown in red in Fig.~\ref{sp_ixpe}.  As a reminder, in the  HEASARC the edge model is the absorption edge, given by 

\[
M(E)=\left\{
\begin{array}{ll}
1 & \text{if } E < E_c \\
\exp\left(-D\left[\frac{E(1+z)}{E_c}\right]^3\right) & \text{if } E > E_c
\end{array}
\right.
\]
where $par1=E_c$ is the threshold energy 1 and $par2=D$ is the absorption depth at the threshold and z is the redshift. 

Taking this spectral model into account, we examined the polarization signal observed at the beginning of the first observation of the {\it IXPE} source in August 2022 using the {\tt polconst} model. The behavior of PD in the initial time period ({\it Epoch 1})  shows an approximately constant PD with  versus energy (see the red dashed line in Fig.~\ref{PD_E}) for all detectors (DU1, DU2, and DU3)  in the {Model 1}. 
{We investigated the difference in fit between two models: constant PD with energy ({\tt polconst}) and increasing PD with energy ({\tt pollin}). We showed that, within the $\chi^2$-test, the {\tt polconst} model better describes the {\it Epoch 1} observations (column 4, orange box in Table~\ref{tab:table_IXPE}) than the {\tt pollin} model (column 5, brown box). Furthermore, the {\it pollin slope} is so small (~0.6\%) that PD can be considered horizontal line in terms of $E(P)$ values (see the blue dashed line in Fig.~\ref{PD_E}). We present all results for both models (Model 1 and {\tt {edge*edge*edge}*tbabs*pollin(Comptb + Gauss)} -- hereinafter Model 2) in Table~\ref{tab:table_IXPE}. 

We also investigated the effect of low photon statistics (observed in Epoch 1) on the conclusion that the polarization degree PD is independent of photon energy. Specifically, we examined another observation segment ({\it Epoch 2}, characterized by a higher number of counts per bin) with the same exposure time as {\it Epoch 1} to observe how PD varies with energy $P(E)$ over the segment {\it Epoch 2}.  During Epoch 2, the IXPE instrument measured the energy-averaged linear polarization dispersion (PD) in the 2--8 keV range of 7.51\% $\pm$ 0.03\% (for polconst) and 7.2\% $\pm$ 0.9\% (for pollin) at a polarization angle (PA) of {18.3}$^{\circ}\pm${1.3}$^{\circ}$ (for polconst) and 15$^{\circ}\pm$1$^{\circ}$ (for pollin) with a statistical confidence of better than 3$\sigma$. Observations in Epoch 2 showed PD that was 2\% higher compared to Epoch 1. Figure~\ref{PD_E} shows the time-averaged polarization characteristic in both Epochs in 4 energy ranges. It is evident from the figure that in Epoch 1 the degree of polarization is localized at a single value, while in Epoch 2 it varies. The polarization angle remains constant with an accuracy of $<3\sigma$ during the observations (Epoch 1 and Epoch 2). Table~\ref{tab:table_IXPE} provides a summary of the measured PD and PA values for various epochs and models.

A comparison of the results showed that the {\it Epoch 1} data are better described by the Model 1 ($\chi^2_{red}$=0.96 for 1301 dof, column 4, orange box) than by the Model 2 ($\chi^2_{red}$=1.32 for 1301 dof, column 5, brown box), while the {\it Epoch 2} data are better described by the Model 2 (column 7, brown box, $\chi^2_{red}$=0.98 for 1341 dof) than by the Model 1 (column 6, orange box, $\chi^2_{red}$=1.25 for 1341 dof). Therefore, the low photon statistics in {\it Epoch 1} does not affect the conclusion that the degree of polarization PD is independent of photon energy. Apparently, a more important factor influencing the pattern $P(E)$ is the intrinsic activity of the source within the observation time segment. Thus, during the interval {\it Epoch 1}, 4U~1630--47 did not change its spectral state, whereas during the interval {\it Epoch 2}, the object changed its spectral state, which could have affected the pattern $P(E)$.

Finally, we applied the F-test procedure for {\it Epoch 1} and showed no significant energy variation of PD and PA. The same F-test procedure applied to {\it Epoch 2} showed that PD increasing with energy preferred at a significance greater than 99.9\%. 
}

The spectral parameters of the source indicate that 4U~1630--47 is in an {intermediate} state or close to it (see also Fig.~\ref{IXPE_RXTE} for a visual comparison with the spectra of 4U~1630--47 in its evolution from LHS through IS to HSS ). Namely, in the $X1$ observation ({\it Epoch 1}) the spectrum of the source is well reproduced by the radiation Comptonization model with the parameters: $\Gamma =2.60\pm0.05$, $N_{com}=0.13\pm0.02~L_{39}/d^2_{10}$, $kT_e=${8.82}$\pm${0.07} keV, $f=0.28\pm0.06$, $kT_s={1}.01\pm0.03$ keV, $E_{Fe}=6.71\pm0.03$ keV, $\sigma_{Fe}=0.07\pm0.02$ keV, and $N_{Fe}=${0.03}$\pm${0.01} photons/s/cm$^2$ ($\chi^2_{red} = 0.96$ for 1303 d.o.f.). 

{In turn, in the observation ({\it Epoch 2}) the spectrum of the source is well reproduced by the radiation Comptonization model with the parameters: $\Gamma =2.52\pm0.07$, $N_{com}=7.43\pm0.09~L_{39}/d^2_{10}$, $kT_e=8.83\pm0.06$ keV, $f=0.28\pm0.06$, $kT_s=1.01\pm0.03$ keV, $E_{Fe}=6.6\pm0.1$ keV, $\sigma_{Fe}=0.2\pm0.1$ keV, and $N_{Fe}=0.17\pm0.08$ photons/s/cm$^2$ ($\chi^2_{red} = 1.25$ for 1341 d.o.f.).} 
In addition to the 6.7 keV emission line, we detected absorption features in the spectrum of 4U~1630--47 during IXPE observations (in the intermediate state), which we {accompanied} with edges at energies of 2.37$\pm$ 0.05, 4.60$\pm$ 0.07 and 7.1$\pm$ 0.1 keV (see Table~\ref{tab:table_IXPE}).

Interestingly, previous observations of 4U~1630--47 also revealed absorption features in the source using different missions: {{\it RXTE}} \citep{Tomsick05},  {{\it Chandra, Swift}, and {\it Suzaku}} \citep{Tomsick14, Kubota07},  {and {\it XRISM}} \citep{Miller25}.  

Using the {\it IXPE} observations, which are characterized by much worse energy resolution than {\it Chandra}/HETG and {\it  XRISM}/Resolve, we hypothesized that our edge detections may mimic the absorption lines with some bias due to a possible wind in 4U~1630--47.

We  investigated the PD error bars in the {\it Epoch 1} time bin, as presented in Fig.~\ref{fraq_1630}. To determine whether this change could be related to a source spectral transition, we analyzed all available simultaneous NICER data simultaneously with the IXPE X1-obsevation. Unfortunately, NuSTAR observations were not available for the {\it Epoch 1} time interval, so we did not consider them in this analysis. Using NICER and {\it IXPE} data, no significant differences were found between {\it Epoch 1} and later epochs in either spectral or timing characteristics. Continuous spectra can be consistently approximated using a  {\tt CompTB} model. 

The NICER spectra also show absorption features, which can be  due to the presence of a disk wind. We modeled these using {\tt constant*edge*edge*edge*TBabs(compTB + Gauss)}. This possible uncertainty in the nature of the line features may also be related to the very short duration of NICER observations during {\it Epoch 1}. 
The lack of significant spectral and timing differences between the NICER simultaneous with {\it Epoch 1} and the subsequent observations may also be due to the very short exposure of the NICER observation during {\it Epoch 1}.

%
%
\begin{table*}
 \caption{The best-fit spectral parameters of the  IXPE observation of 4U~1630--47 for the spectro-polarimetric {models} {\tt constant*edge*edge*edge*tbabs*polconst(CompTB+Gauss)} {(orange)} and {{\tt constant*edge*edge*edge*tbabs*pollin(CompTB+Gauss)} (brown)}.  From left to right are, (1) model components; (2) parameters in components; (3) {parameter {description}} {{for \it Epoch 1}: with \tt{polconst} (4) and  \tt{pollin} (5) models, as well as for {\it Epoch 2}: with \tt{polconst} (6) and  \tt{pollin} (7) models}. 
}
   \label{tab:table_IXPE}
\resizebox{1\textwidth}{!}{%
 \begin{tabular}{lllllllr}
 \hline\hline  
\multicolumn{1}{|c|}{Model}   &         Model component   &\multicolumn{1}{|c|}{Description}    &  \multicolumn{4}{|c|} {    Parameter value}            \\
\rowcolor[gray]{.9}       &\multicolumn{1}{|c|}{}              &\multicolumn{1}{|c|}{} &  \multicolumn{2}{|c|} {Epoch 1} & \multicolumn{2}{|c} {Epoch 2}\\
 \hline
        &                        Data group: 1 &&  &\\
Edge1   &    E$_1$*,      keV  & Threshold energy & \multicolumn{2}{c} {2.37 $\pm$ 0.05}&\multicolumn{2}{c} {2.37 $\pm$ 0.05}\\
        &   $\tau^{max}_1*$    & Absorption depth at the threshold &\multicolumn{2}{c} {0.07 $\pm$ 0.01}&\multicolumn{2}{c} {0.07 $\pm$ 0.01}\\   
Edge2   &    E$_2$*,      keV  & Threshold energy &\multicolumn{2}{c} {4.60 $\pm$ 0.07}&\multicolumn{2}{c} {4.60 $\pm$ 0.07}\\
        &   $\tau^{max}_2*$    & Absorption depth at the threshold &\multicolumn{2}{c} {0.34 $\pm$ 0.01}&\multicolumn{2}{c}{0.34 $\pm$ 0.01}\\
Edge3   &    E$_3$*,       keV & Threshold energy &\multicolumn{2}{c}{7.1 $\pm$ 0.1}&\multicolumn{2}{c}{7.1 $\pm$ 0.1}\\
        &    $\tau^{max}_3*$   & Absorption depth at the threshold &\multicolumn{2}{c}{1.72$\pm$ 0.03}&\multicolumn{2}{c}{1.72$\pm$ 0.03}\\   
TBabs    &  $N_H^{(1)}$,     10$^{22}$cm$^{-2}$  & Equivalent hydrogen column &\multicolumn{2}{c}{6.13 $\pm$ 0.05}&\multicolumn{2}{c}{6.13 $\pm$ 0.05}\\ 
\rowcolor[rgb]{0.9290 0.6940 0.1250}Constant* (for $polconst$)  &         & Energy-independent constant   factor& 0.92 $\pm$ 0.09&\cellcolor{white}...&1.67 $\pm$ 0.08&\cellcolor{white}...\\    
\rowcolor[rgb]{0.9290 0.6940 0.1250}polconst &  A$_1^{polconst}$*, \%  & Polarization fraction &    5.8 $\pm$ 0.7&\cellcolor{white}... & 7.51 $\pm$ 0.03&\cellcolor{white}...\\
\rowcolor[rgb]{0.9290 0.6940 0.1250}         &  $\Psi_1^{polconst}$*,        deg & Polarization angle &     18.3 $\pm$ 1.2&\cellcolor{white}...&18.3 $\pm$ 1.3&\cellcolor{white}...\\
\rowcolor[rgb]{0.8500 0.3250 0.0980}Constant* (for $pollin$)  &            & Energy-independent constant   factor &\cellcolor{white}...& 1.03 $\pm$ 0.07&\cellcolor{white}...& 1.07 $\pm$ 0.09\\    
\rowcolor[rgb]{0.8500 0.3250 0.0980}pollin &  A$_1^{pollin}$*, \%            & Polarization fraction at 1 keV &\cellcolor{white}...& 7.4 $\pm$ 0.6 &\cellcolor{white}...& 7.2 $\pm$ 0.9\\
\rowcolor[rgb]{0.8500 0.3250 0.0980}         &  A$_1^{slope}*$,  \%  & Polarization fraction slope &\cellcolor{white}...&0.6 $\pm$ 0.1 &\cellcolor{white}... & 
{1.36} $\pm$ 
{0.12}\\
\rowcolor[rgb]{0.8500 0.3250 0.0980}         &  $\Psi_1^{pollin}$*,       deg & Polarization angle at 1 keV &\cellcolor{white}... & 14.5$\pm$ 1.3 & \cellcolor{white}... & 15.3 $\pm$ 1.3\\
\rowcolor[rgb]{0.8500 0.3250 0.0980}         &  $\Psi_1^{slope}$*, deg & Polarization angle slope &\cellcolor{white}... & 4.9 $\pm$ 1.7 & \cellcolor{white}... & 4.7 $\pm$ 1.1\\
Comptb   &   $kT_s^{(1)}$,        keV & Temperature of the seed photons &\multicolumn{2}{c}{1.01 $\pm$ 0.03}&\multicolumn{2}{c}{1.01 $\pm$ 0.03} \\
         &  $\alpha^{(1)}$   & Energy index of the Comptonization spectrum &\multicolumn{2}{c}{1.60 $\pm$ 0.05}&\multicolumn{2}{c}{1.52 $\pm$ 0.07}\\
         & $kT_e^{(1)}$,        keV & Temperature of the electrons &\multicolumn{2}{c}{8.82 $\pm$ 0.07}&\multicolumn{2}{c}{8.83 $\pm$ 0.06}\\
         & log(A)*  & Log of the illuminating factor parameter &\multicolumn{2}{c}{-0.44 $\pm$ 0.06}&\multicolumn{2}{c}{-0.39 $\pm$ 0.04}\\
         & $N_{com}^{(1)\dagger}$    & Normalization of the seed photon spectrum &\multicolumn{2}{c}{0.13 $\pm$ 0.02}&\multicolumn{2}{c}{7.43 $\pm$ 0.09}\\
Gaussian  & $E_{line}^{(1)}$,      keV & Line energy &\multicolumn{2}{c}{6.71 $\pm$ 0.03}&\multicolumn{2}{c}{6.6 $\pm$ 0.1} \\
         & $N_{line}^{(1)\dagger\dagger}$  &Line normalization &\multicolumn{2}{c}{0.03 $\pm$ 0.01}&\multicolumn{2}{c}{0.17 $\pm$ 0.08}\\
 \hline
                   &        Data group: 2& & & \\
Edge1   &    E$_1$,      keV  & Threshold energy & \multicolumn{2}{c}{2.37=E$_1$*$^{~(fixed)}$}&\multicolumn{2}{c} {2.37=E$_1$*$^{~(fixed)}$}\\
        &   $\tau^{max}_1$    & Absorption depth at the threshold &\multicolumn{2}{c} {0.07=$\tau_1*^{max~(fixed)}$}&\multicolumn{2}{c} {0.07=$\tau_1*^{max~(fixed)}$}\\   
Edge2   &    E$_2$,      keV  & Threshold energy &\multicolumn{2}{c} {4.60=E$_2$*$^{~(fixed)}$}&\multicolumn{2}{c} {4.60=E$_2$*$^{~(fixed)}$}\\
        &   $\tau^{max}_2$    & Absorption depth at the threshold &\multicolumn{2}{c} {0.34=$\tau_2*^{max~(fixed)}$}&\multicolumn{2}{c} {0.34=$\tau_2*^{max~(fixed)}$}\\
Edge3   &    E$_3$,       keV & Threshold energy &\multicolumn{2}{c}{7.1=E$_3$*$^{~(fixed)}$}&\multicolumn{2}{c}{7.1=E$_3$*$^{~(fixed)}$}\\
        &    $\tau^{max}_3$   & Absorption depth at the threshold &\multicolumn{2}{c}{1.72=$\tau_3*^{max~(fixed)}$}&\multicolumn{2}{c}{1.72=$\tau_3*^{max~(fixed)}$}\\   
TBabs    &  $N_H^{(2)} $,     10$^{22}$cm$^{-2}$  & Equivalent hydrogen column &\multicolumn{2}{c}{6.57$\pm$0.05}&\multicolumn{2}{c}{6.57$\pm$0.05}\\ 
\rowcolor[rgb]{0.9290 0.6940 0.1250}polconst &  A$_2^{polconst}$, \%  & Polarization fraction &    5.8=A$_1^{polconst}$*&\cellcolor{white}... & 7.51=A$_1^{polconst}$*&\cellcolor{white}...\\
\rowcolor[rgb]{0.9290 0.6940 0.1250}         &  $\Psi_2^{polconst}$*,        deg & Polarization angle &     18.3=$\Psi_1^{polconst}$*&\cellcolor{white}...&18.3=$\Psi_1^{polconst}$*&\cellcolor{white}...\\
\rowcolor[rgb]{0.8500 0.3250 0.0980}pollin &  A$_2^{pollin}$, \%            & Polarization fraction at 1 keV &\cellcolor{white}...&7.4=A$_1^{pollin}$*&\cellcolor{white}...&7.2=A$_1^{pollin}$*\\
\rowcolor[rgb]{0.8500 0.3250 0.0980}         &  A$_2^{slope}$,  \%  & Polarization fraction slope & \cellcolor{white}...&0.6=A$_1^{slope}*$&\cellcolor{white}...&{1.36}=A$_1^{slope}*$\\
\rowcolor[rgb]{0.8500 0.3250 0.0980}         &  $\Psi_2^{pollin}$,       deg & Polarization angle at 1 keV &\cellcolor{white}...&14.5=$\Psi_1^{pollin}$*&\cellcolor{white}...&15.3=$\Psi_1^{pollin}$*\\
\rowcolor[rgb]{0.8500 0.3250 0.0980}         &  $\Psi_2^{slope}$, deg & Polarization angle slope &\cellcolor{white}...&4.9=$\Psi_1^{slope}$*&\cellcolor{white}...&4.7=$\Psi_1^{slope}$*\\
Comptb   &   $kT_s$,        keV & Temperature of the seed photons &\multicolumn{2}{c}{0.96 $\pm$ 0.05}&\multicolumn{2}{c}{0.91 $\pm$ 0.01} \\
         &  $\alpha^{(2)}$   &Energy index of the Comptonization spectrum &\multicolumn{2}{c}{1.52 $\pm$ 0.07}&\multicolumn{2}{c}{1.48 $\pm$ 0.09}\\
         & $kT_e^{(2)}$,        keV & Temperature of the electrons &\multicolumn{2}{c}{8.80 $\pm$ 0.09}&\multicolumn{2}{c}{8.82 $\pm$ 0.07}\\
         & $N_{com}^{(2)\dagger}$    & Normalization of the seed photon spectrum &\multicolumn{2}{c}{0.13=$N_{com}^{(1)}$}&\multicolumn{2}{c}{7.43=$N_{com}^{(1)}$}\\
Gaussian  & $E_{line}^{(2)}$,      keV & Line energy &\multicolumn{2}{c}{6.86 $\pm$ 0.07}&\multicolumn{2}{c}{6.7 $\pm$ 0.1} \\
 \hline
 &                           Data group: 3 & &&\\
Edge1   &    E$_1$,      keV  & Threshold energy & \multicolumn{2}{c} {2.37=E$_1$*$^{~(fixed)}$}&\multicolumn{2}{c} {2.37=E$_1$*$^{~(fixed)}$}\\
        &   $\tau^{max}_1$    & Absorption depth at the threshold &\multicolumn{2}{c}{0.07=$\tau_1*^{max~(fixed)}$}&\multicolumn{2}{c}{0.07=$\tau_1*^{max~(fixed)}$}\\
Edge2   &    E$_2$,      keV  & Threshold energy &\multicolumn{2}{c} {4.60=E$_2$*$^{~(fixed)}$}&\multicolumn{2}{c} {4.60=E$_2$*$^{~(fixed)}$}\\
        &   $\tau^{max}_2$    & Absorption depth at the threshold &\multicolumn{2}{c} {0.34=$\tau_2*^{max~(fixed)}$}&\multicolumn{2}{c} {0.34=$\tau_2*^{max~(fixed)}$}\\
Edge3   &    E$_3$,       keV & Threshold energy &\multicolumn{2}{c}{7.1=E$_3$*$^{~(fixed)}$}&\multicolumn{2}{c}{7.1=E$_3$*$^{~(fixed)}$}\\
        &    $\tau^{max}_3$   & Absorption depth at the threshold &\multicolumn{2}{c}{1.72=$\tau_3*^{max~(fixed)}$}&\multicolumn{2}{c}{1.72=$\tau_3*^{max~(fixed)}$}\\   
TBabs    &  $N_H^{(3)} $,     10$^{22}$cm$^{-2}$  & Equivalent hydrogen column &\multicolumn{2}{c}{6.9$\pm$0.1}&\multicolumn{2}{c}{6.57$\pm$0.05}\\ 
\rowcolor[rgb]{0.9290 0.6940 0.1250}polconst &  A$_3^{polconst}$, \%  & Polarization fraction &    5.8=A$_1^{polconst}$*&\cellcolor{white}... & 7.51=A$_1^{polconst}$*&\cellcolor{white}...\\
\rowcolor[rgb]{0.9290 0.6940 0.1250}         &  $\Psi_3^{polconst}$,        deg & Polarization angle &     18.3=$\Psi_2^{polconst}$*&\cellcolor{white}...&18.3=$\Psi_2^{polconst}$*&\cellcolor{white}...\\
\rowcolor[rgb]{0.8500 0.3250 0.0980}pollin &  A$_3^{pollin}$, \%            & Polarization fraction at 1 keV &\cellcolor{white}...&7.4=A$_1^{pollin}$*&\cellcolor{white}...&7.2=A$_1^{pollin}$*\\
\rowcolor[rgb]{0.8500 0.3250 0.0980}         &  A$_3^{slope}$,  \%  & Polarization fraction slope &\cellcolor{white}...&0.6=A$_1^{slope}$*&\cellcolor{white}...&
{1.36}=A$_1^{slope}$*\\
\rowcolor[rgb]{0.8500 0.3250 0.0980}         &  $\Psi_3^{pollin}$,       deg & Polarization angle at 1 keV &\cellcolor{white}...&14.5=$\Psi_1^{pollin}$* &\cellcolor{white}...&15.3=$\Psi_1^{pollin}$*\\
\rowcolor[rgb]{0.8500 0.3250 0.0980}         &  $\Psi_3^{slope}$, deg & Polarization angle slope &\cellcolor{white}...&4.9=$\Psi_1^{slope}$*&\cellcolor{white}...&4.7=$\Psi_1^{slope}$*\\
Comptb   &   $kT_s^{(3)}$,        keV & Temperature of the seed photons &\multicolumn{2}{c}{0.93 $\pm$ 0.06}&\multicolumn{2}{c}{0.88 $\pm$0.01} \\
         &  $\alpha^{(3)}$   &Energy index of the Comptonization spectrum &\multicolumn{2}{c}{1.36 $\pm$ 0.03}&\multicolumn{2}{c}{1.51 $\pm$ 0.08}\\
         & $kT_e^{(3)}$        keV & Temperature of the electrons &\multicolumn{2}{c}{8.80 $\pm$ 0.06}&\multicolumn{2}{c}{8.83 $\pm$ 0.06}\\
         & $N_{com}^{(3)\dagger}$    & Normalization of the seed photon spectrum &\multicolumn{2}{c}{0.13=$N_{com}^{(1)}$}&\multicolumn{2}{c}{7.43=$N_{com}^{(1)}$}\\
Gaussian  & $E_{line}^{(3)}$,      keV & Line energy &\multicolumn{2}{c}{6.61 $\pm$ 0.04}&\multicolumn{2}{c}{6.7 $\pm$ 0.2} \\
\hline
   & $\chi^2_{red}$ (dof) &&\cellcolor[rgb]{0.9290,0.6940,0.1250}0.96 (1303)& \cellcolor[rgb]{0.8500 0.3250 0.0980}1.32 (1303)&\cellcolor[rgb]{0.9290,0.6940,0.1250}1.25 (1341)&\cellcolor[rgb]{0.8500 0.3250 0.0980}0.98 (1341)\\
      \hline
      \hline
 \end{tabular}
}
\\
{* These parameters obtained for DU1 and fixed for DU2 and DU3;} 
$^{\dagger}$ The {\tt  CompTB} component {normalization} is in units of $L_{39}/d_{10}$, where $L_{39}$ is the source luminosity in units of  $10^{39}$ erg/s and $d_{10}$ is the distance to the source in units of 10 kpc.  
$^{\dagger\dagger}$ {\tt Gaussian} component is in units of total photons cm$^{-2}$s$^{-1}$ in line. 
\end{table*}
 
Since the spectral best-fit parameters, as well as a comparison of the shape and position of the spectrum of 4U~1630--47 during IXPE observations (see red histogram in Fig.~\ref{IXPE_RXTE}) with the source spectra previously observed using RXTE (Fig.~\ref{IXPE_RXTE}), indicate that the source is in {an intermediate} state,  we established the independence of the polarization degree  ($\sim$ {6}\%) on the photon energy in the {IS} state for 4U 1630--47. This indicates that testing the polarization properties of 4U~1630--47 and other XRBs for its PD energy independence should be carried out only when the source is in a single spectral state. This is entirely consistent with the conclusions presented in ST85.

\begin{figure}
\centering
\includegraphics[scale=0.62,angle=0]{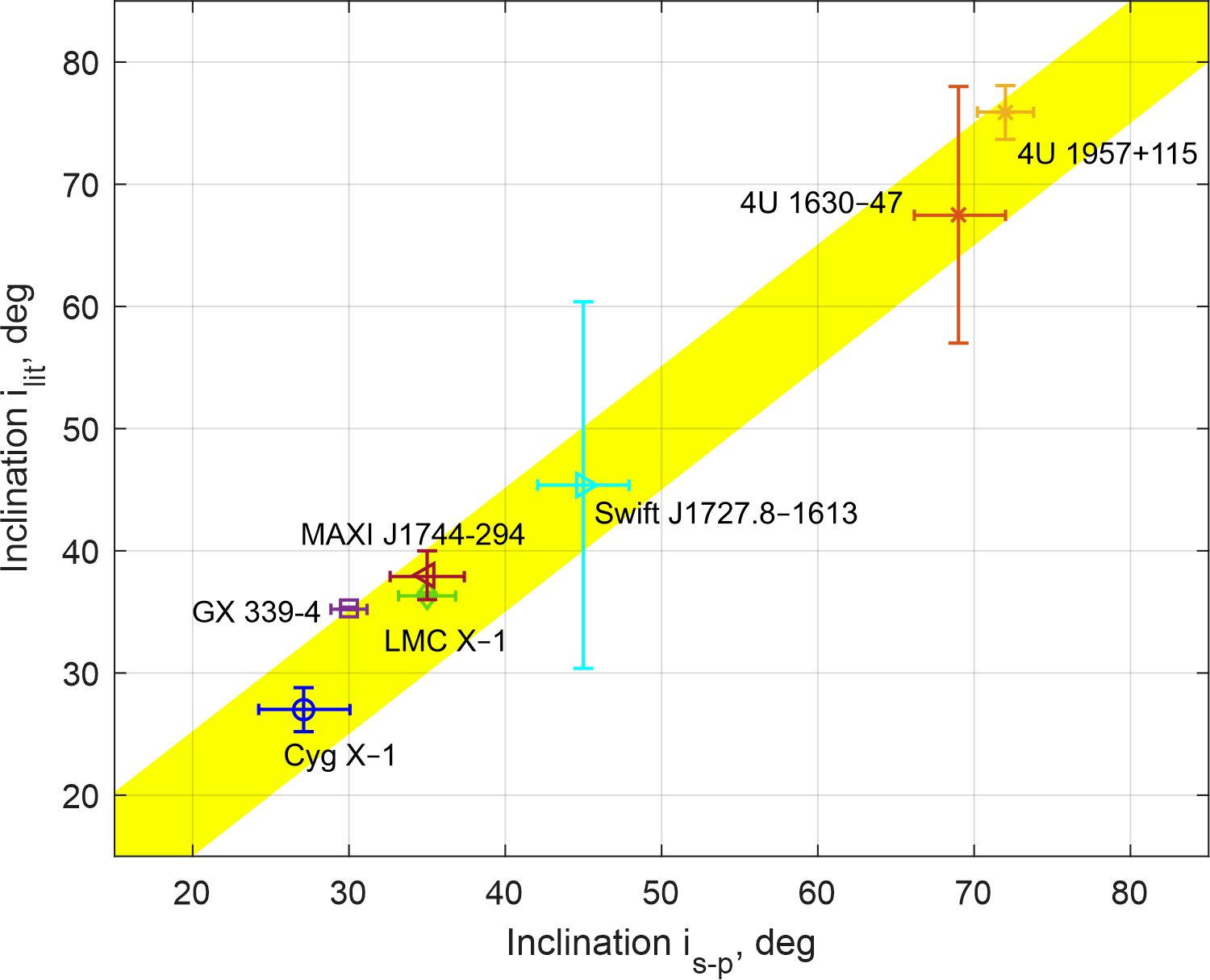} 
\caption{Comparison of the spectropolarimetrically determined inclinations $i_{sp-pol}$ (using the $\alpha-\tau-T_e$ dependence from ST85) with the corresponding inclinations obtained by optical or spectral fitting methods found in the literature, $i_{lit}$, for XRBs Cyg~X--1, 4U~1630--47, 4U~1957+115, Cyg~X--3, LMC~X--1, Swift~J1727.8--1613, and MAXI~J1744--294. Data are taken from \cite{Tetal25,TSSM25,SE26}. 
The yellow strip indicates an inclination spread of 5 degrees. 
 }
\label{inck-compar_sp-pol-lit}
\end{figure}

We plotted the obtained polarization degree $P$ for 4U~1630--47 on the $P-\mu-\tau$ diagram (taken from ST85), where $\mu=\cos(i)$ and $i$ is the disk inclination angle to the line of sight of an Earth observer  (Fig.~\ref{pol_vs_mu}).  To estimate the inclination for 4U~1630--47  from the known polarization degree \citep{RodriguezCavero23,Rawat23,Ratheesh24} using this diagram, we should use information on the optical depth $\tau_0$, which can be obtained from spectral analysis (see Section~\ref{spectral analysis}). Indeed, the spectra of the X-ray source are well described by the Comptonization model with a photon index $\Gamma$ of approximately 2.6 and a plasma temperature $kT_e \sim 8$ keV. From here, it is easy to estimate the half-optical depth of the CC $\tau_0\sim 1.8$ (see \cite{ST85, Tetal25, TSSM25}). Using the Tetal25 method, we can estimate the inclination as $
{69\pm1}^{\circ}$, which is consistent with previous estimates (e.g., \cite{Kuulkers98, STS16, RodriguezCavero23}). Note that this method, now applied for 4U~1630--47, has already been successfully tested on a number of XRBs (e.g., \cite{SE26,Tetal25,TSSM25}, see Fig.~\ref{inck-compar_sp-pol-lit}). In Fig.~\ref{pol_vs_mu}, we also plotted points for other known BH XRBs in different spectral states to demonstrate the performance of the Tetal25 method. Thus, for all the presented XRBs, the parameters $P$, $\mu$, and $\tau_0$ fit this diagram well, confirming the acceptability of the ST85 solution for the ``flat pancake'' case and indicating the presence of flat accretion disks (at least flat inner CC regions of the disk) in these XRBs.

Note that the null hypothesis probability value of 0.029 ($\sim$ 3\%) obtained by fitting the {\it IXPE} spectral data at {\it Epoch 1} indicates that the observed fluctuations occurred by chance is only 3\%. Assuming the null hypothesis is true, this value is below the standard threshold of 5\% (0.05). Thus, these results are considered statistically significant, rejecting the null hypothesis and confirming the existence of a real effect.

%
%

\section{Discussion  and conclusions }
\label{sect:results_discussion}

Previously, \cite{Kushwaha23} argued that the polarization degree, PD depends on the photon energy in 4U~1630--47, using long-term IXPE observations of this source, where the X-ray source exhibited multiple transitions from one to another spectral state. They claimed that the PD is {\it an increasing function of energy} which is opposite  to  that  ST85 predicted   that the PD is independently of the photon energy. In fact, the X-ray photons are results of the up-scattering of the disk (soft) photons  and their polarization dependence  is  smeared out  (see  Fig.~\ref{pol_vs_mu}) and the PD  value is determined  by the optical depth of the flat CC [see the proof of this statement in ST85 and Tetal25{]}. 

\begin{figure}
\centering
\includegraphics[scale=0.68,angle=0]{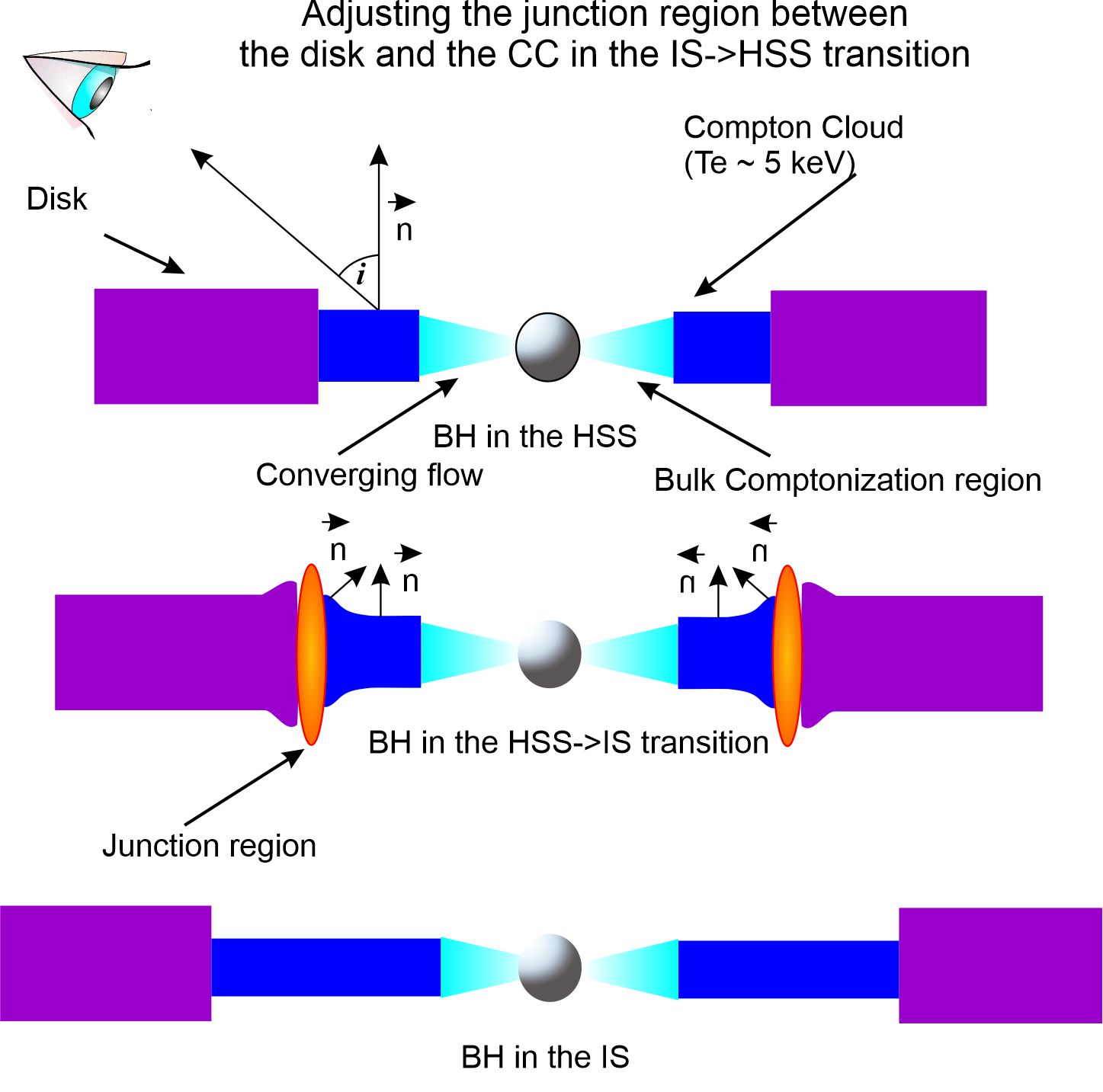} 
\caption{
{Rearrangement between the inner edge of the disk and the outer edge of the CC during the transition from the LHS to the IS. New boundary conditions arise at the boundary between these regions (orange zone), accompanied by the formation of shock waves, which leads to a curvature of the CC shape in the zone adjacent to this boundary. X-ray emission generated in this zone will be polarized in the curved (rather than flat) layer.}
 }
\label{model+boudary}
\end{figure}

{To test our energy independence of the polarization degree, $P(E)$, compared to the results of} \cite{Kushwaha23, Ratheesh24}, {who considered all segments, we excluded the effect of low photon statistics (Sect.~\ref{spectral analysis_ixpe}) and analyzed the physical processes that explain why the polarization behavior at Epoch 1 differs from its behavior at other times. We found large uncertainties in the parameters at other epochs (besides Epoch 1), caused, for example, by frequent changes in spectral states. For example, at Epoch 1, the object 4U~1630--47 is in a single spectral state (intermediate), while at other epochs it exhibits a ``mixture'' of different spectral states, making it impossible to conduct purely energy-dependent polarization diagnostics free of transition effects. This is due to the fact that during the transition between states (e.g., from LHS to IS), restructuring occurs between the inner edge of the disk and the outer edge of the CC. At the boundary between these regions, new boundary conditions arise, accompanied by the formation of shock waves (a jump in the derivatives at the layer boundary when solving the transport equation; see, e.g., }\cite{TK23}) {and, ultimately, a curvature of the CC shape in the zone adjacent to this boundary (Fig.~\ref{model+boudary}). The radiation generated in this zone will be polarized in the curved (rather than flat) layer. This will contribute to the overall polarization, which will result from polarization in both the flat and curved layers.} 

{In addition, the $P(E)$ pattern may be influenced by relativistic effects. In fact, relativistic effects increase the degree of polarization P with increasing energy E (in 2--8 keV range) due to general relativity effects near black holes} (\cite{Krawczynski24}).

%
%

 We present the results of spectral and polarimetric analysis of X-ray outburst episodes from 4U~1630--47 observed by {\it IXPE} and {\it NICER}. We argued that the X-ray source broadband energy spectra during all 
states can be reproduced by a physical model composed 
of the Comptonized component ({\tt CompTB}), due to the presence of the transition layer (or CC), and an {\it Gaussian} iron-line component. We investigated the time interval of the {\it IXPE} observations during which 4U~1630--47 was strictly in the  {intermediate} state. Spectral analysis using this model shows that during the {intermediate} state, the X-ray source 4U~1630--47 was characterized by a Comptonized photon index $\Gamma\sim 2.5$ and an electron temperature $kT_e\sim ~9 ~{\rm keV}$.  During  this {intermediate} state in 4U~1630--47, we detected a quasi-constant degree of linear polarization $P$ of the X-ray source emission, {independent} on the photon energy, as predicted in ST85. 
We take  this constancy of $P(E)$ as a result of Comptonization (upward scattering) of soft photons in a flat CC. This information allowed us to determine  of half-optical depth of the CC    $\tau_0\sim 1.8$ and the inclination ${69\pm1}^\circ$ in 4U~1630--47 using the $(P,\cos(i),\tau_0$) diagram predicted by ST85. This inclination value is close to that  value estimated by other methods: scaling method gives us ($\sim$ 70$^\circ$) and dynamical (60$^\circ$--75$^\circ$).  Note that we used the $(P,\cos(i),\tau_0$)  diagram taking into account positive values of the polarization degree $P$. In fact, one can argue that due to uncertainty in the determination of the observational sign of the polarization, the linear polarization $P$ can be positive. But in this case the resulting optical depth of half of the CC $\tau_0$ is much greater than 1 (see Fig.~\ref{pol_vs_mu}) and thus the observational spectrum is the Wien spectrum, which is not our case.  Negative values of $P$ imply (see ST85) an oscillation of the electric field in the meridional plane (the plane defined by the normal to the disk and the direction of propagation). This is precisely what is observed by the {\it IXPE} in X-ray BH binaries and Seyfert 1 AGNs with spatially resolved radio jet emission \citep{Tetal25}.

\begin{acknowledgements}
The Imaging X-ray Polarimetry Explorer (IXPE) is a joint mission of the US and Italy.  The US contribution is supported by the National Aeronautics and Space Administration (NASA) and led and managed by its Marshall Space Flight Center (MSFC), with industry partner Ball Aerospace (now, BAE Systems).  The Italian contribution is supported by the Italian Space Agency (Agenzia Spaziale Italiana, ASI) through contract ASI-OHBI-2022-13-I.0, agreements ASI-INAF-2022-19-HH.0 and ASI-INFN-2017.13-H0, and its Space Science Data Center (SSDC) with agreements ASI-INAF-2022-14-HH.0 and ASI-INFN 2021-43-HH.0, and by the Istituto Nazionale di Astrofisica (INAF) and the Istituto Nazionale di Fisica Nucleare (INFN) in Italy.  This research used data products provided by the IXPE Team (MSFC, SSDC, INAF, and INFN) and distributed with additional software tools by the High-Energy Astrophysics Science Archive Research Center (HEASARC), at NASA Goddard Space Flight Center (GSFC). { We acknowledge the reviewers and editor for their in-depth analysis of the paper 
and useful comments that improved our paper. } The study was conducted under the state assignment of Lomonosov Moscow State University. 
\end{acknowledgements}


\end{document}